\documentstyle[12pt,psfig]{article}

\renewcommand{\textfraction}{0.2}

\begin{document}

\begin{titlepage}
\begin{center}
\today     \hfill    LBNL-38911 \\
~{} \hfill UCB-PTH-96/22  \\

\vskip .25in

{\large \bf Supersymmetric Framework for\\ a Dynamical Fermion 
Mass Hierarchy}\footnote{This 
work was supported in part by the Director, Office of 
Energy Research, Office of High Energy and Nuclear Physics, Division of 
High Energy Physics of the U.S. Department of Energy under Contract 
DE-AC03-76SF00098 and in part by the National Science Foundation under 
grant PHY-95-14797.}

\vskip 0.3in

Nima Arkani-Hamed$^{1,2}$, Christopher D. Carone$^1$, 
Lawrence J. Hall$^{1,2}$, and Hitoshi Murayama$^{1,2}$

\vskip 0.1in

{{}$^1$ \em Theoretical Physics Group\\
     Lawrence Berkeley National Laboratory\\
     University of California, Berkeley, California 94720}

\vskip 0.1in

{{}$^2$ \em Department of Physics\\
     University of California, Berkeley, California 94720}

\end{center}

\vskip .1in

\begin{abstract}
We propose a new framework for constructing supersymmetric theories
of flavor, in which flavor symmetry breaking is triggered by
the dynamical breakdown of supersymmetry at low energies.  All mass scales 
in our scheme are generated from the supersymmetry breaking scale
$\Lambda_{SSB} \approx 10^7$ GeV through radiative corrections.  We assume
a spontaneously broken flavor symmetry and the Froggatt-Nielsen mechanism
for generating the fermion Yukawa couplings. Supersymmetry breaking radiatively
induces a vacuum expectation value for a scalar field, which generates 
invariant masses for the Froggatt-Nielsen fields at $M_F \approx 10^4$ GeV.  
`Flavon' fields $\varphi$, which spontaneously break the flavor symmetry, 
naturally acquire negative squared masses due to two-loop diagrams involving 
the Froggatt-Nielsen fields, and acquire vacuum expectation values of order
$\langle \varphi \rangle \approx M_F/16\pi^2$.  The fermion mass hierarchy 
arises in our framework as a power series in the ratio 
$\langle \varphi \rangle / M_F \approx 1/16\pi^2$.  
\end{abstract}

\end{titlepage}
\renewcommand{\thepage}{\roman{page}}
\setcounter{page}{2}
\mbox{ }

\vskip 1in

\begin{center}
{\bf Disclaimer}
\end{center}

\vskip .2in

\begin{scriptsize}
\begin{quotation}
This document was prepared as an account of work sponsored by the United
States Government. While this document is believed to contain correct 
information, neither the United States Government nor any agency
thereof, nor The Regents of the University of California, nor any of their
employees, makes any warranty, express or implied, or assumes any legal
liability or responsibility for the accuracy, completeness, or usefulness
of any information, apparatus, product, or process disclosed, or represents
that its use would not infringe privately owned rights.  Reference herein
to any specific commercial products process, or service by its trade name,
trademark, manufacturer, or otherwise, does not necessarily constitute or
imply its endorsement, recommendation, or favoring by the United States
Government or any agency thereof, or The Regents of the University of
California.  The views and opinions of authors expressed herein do not
necessarily state or reflect those of the United States Government or any
agency thereof, or The Regents of the University of California.
\end{quotation}
\end{scriptsize}

\vskip 2in

\begin{center}
\begin{small}
{\it Lawrence Berkeley Laboratory is an equal opportunity employer.}
\end{small}
\end{center}

\newpage
\renewcommand{\thepage}{\arabic{page}}
\setcounter{page}{1}
\section{Introduction} \label{sec:intro}

Two outstanding problems of particle physics both involve the origin of
symmetry breaking. How does the electroweak gauge symmetry break, allowing the
$W$ and $Z$ bosons to acquire mass? And secondly, what breaks the flavor symmetry
of the standard model gauge interactions, allowing the quark and leptons to
also become massive? The mechanisms for these symmetry breakings must involve
new particles and interactions. Furthermore, this new physics must involve new
mass scales: the physics of electroweak symmetry breaking (EWSB) must provide
an origin for the weak scale, $M_Z$, and the physics of flavor symmetry
breaking (FSB) must involve a mass scale, $M_F$.

In the standard model, the interactions of the Higgs doublet, 
$H$,  generate both EWSB and FSB. 
Although the Higgs sector is extremely it provides no
understanding for the small size of the weak scale, 
$\langle H \rangle/M_{Pl}$, nor for
the small size of FSB, $m_{q,l}/M_Z$. Indeed, the Yukawa couplings of the
standard model are arbitrary, explicit, FSB parameters. An understanding of
fermion masses would result if these small dimensionless parameters were given
in terms of a small ratio $\langle \varphi \rangle /M_F$ \cite{FN}, where $\langle 
\varphi \rangle $ is the vacuum expectation value (vev) of a flavon field
which spontaneously breaks a flavor group $G_f$. However, such a scheme
involves three mass scales: $\langle H \rangle, \langle \varphi \rangle$ and
$M_F$. 

In theories with weak scale supersymmetry, the weak scale, 
$\langle H \rangle$, is determined to be comparable to the superpartner 
masses, $\tilde{m}$, which are derived from two other scales: the 
primordial supersymmetry breaking scale $\Lambda_{SSB}$ and the messenger 
scale $M_{mess}$. Specific models show that it is possible to generate 
supersymmetry breaking, and therefore $\Lambda_{SSB}$, by dimensional
transmutation from non-perturbative dynamics \cite{ADS2}. The messenger scale
describes the softness of the superpartner masses, $\tilde{m}$, which 
rapidly vanish at scales above $M_{mess}$. 

Hence in supersymmetric theories there are generally four mass scales: two to
describe supersymmetry breaking $\Lambda_{SSB}$ and $M_{mess}$, which lead to
EWSB, and two to describe flavor physics and FSB, $M_F$ and
$\langle \varphi \rangle $. In supergravity theories, supersymmetry breaking 
is transmitted to superpartners via supergravitational interactions, so 
that $M_{mess} = M_{Pl}$ and $\Lambda_{SSB}$ is determined to be
$10^{11}$ GeV. If $\Lambda_{SSB} < 10^{11}$ GeV, sufficient supersymmetry
breaking can be transmitted to the superpartners by gauge interactions, and it
is this case which we study in this paper \cite{ACW,DF}. In these theories
$M_{mess} \approx 1/(16 \pi^2) \Lambda_{SSB}$ can arise in perturbation theory
\cite{DN}.  The messenger sector contains a set of vector-like generations,
$X$ and $\bar{X}$, which acquire both supersymmetry preserving and
supersymmetry breaking masses: $M_{mess} [\bar{X} X]_F + M_{mess}^2
[\bar{X} X]_A$.  On integrating these heavy vector generations out of the
theory, the standard model gauge interactions transmit the supersymmetry
breaking to the superpartners, giving $\tilde{m} \approx (1/16 \pi^2)^2
\Lambda_{SSB}$.  Furthermore, renormalization group scalings induced by the
large top quark Yukawa coupling, $\lambda_t$, induce a negative shift in the
Higgs mass squared: $\Delta m_H^2/ m_H^2 \approx -3/4 \pi^2 \ln(M_{mess}/300
\mbox{ GeV}) (m^2_{\tilde{t}}/m_H^2)$. Since $m_{\tilde{t}}/m_H \approx
\alpha_3/\alpha_2$, this triggers EWSB. Thus $M_{mess}$ and $\langle H
\rangle$ are understood as arising from $\Lambda_{SSB}$ by a successive
cascade of perturbative loop factors:
\begin{equation}
\Lambda_{SSB} \rightarrow M_{mess} \rightarrow \tilde{m}, \langle H \rangle
\label{eq:EWSB}
\end{equation}
In this paper we study whether the scales of FSB can be similarly
derived from $\Lambda_{SSB}$ by a succession of perturbative loops:
\begin{equation}
\Lambda_{SSB} \rightarrow M_F \rightarrow \langle \varphi \rangle
\label{eq:FSB}
\end{equation}

\begin{figure}[t]
	\centerline{\psfig{file=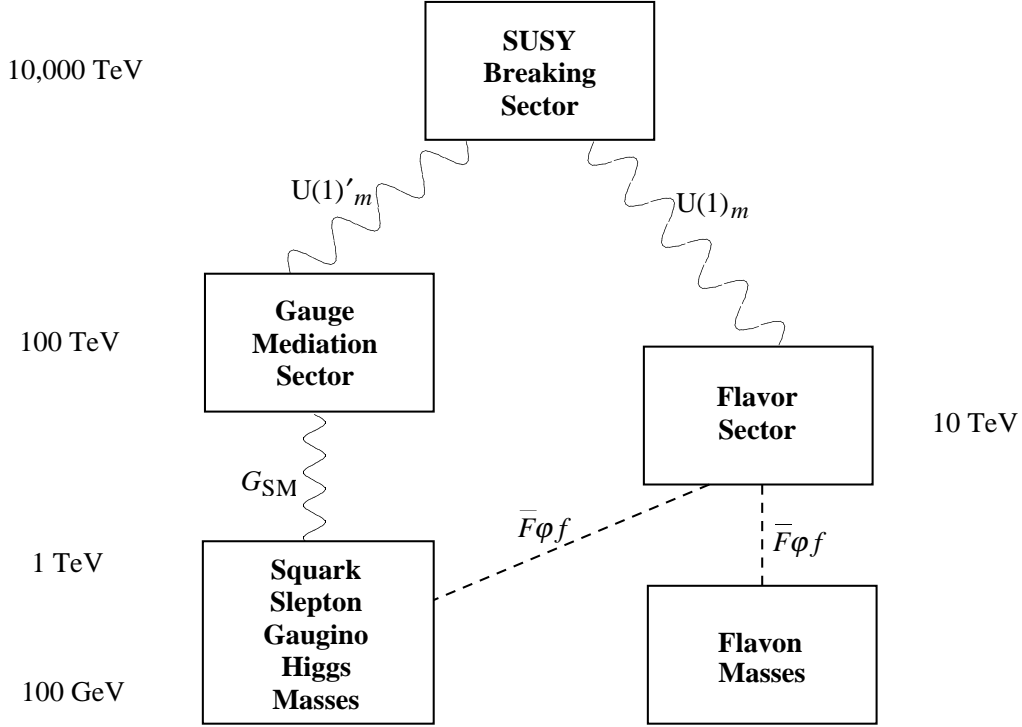,width=\textwidth}}
	\caption{Schematic structure of the model.  The wavy lines indicate 
	radiative corrections due to gauge interactions, while the dashed 
	lines due to superpotential interactions.}
	\protect\label{schematic}
\end{figure}

The most successful scheme for generating fermion mass hierarchies from flavor
symmetry breaking vevs, $\langle \varphi \rangle$, involves the mixing of 
heavy vector-like generations, $F$ and $\bar{F}$, with the light generations, 
$f$: $[M_F \bar{F} F + \bar{F} \varphi f]_F$. The flavor symmetry group,
$G_f$, 
prevents direct Yukawa couplings for the light quarks and leptons $[ffH]_F$, 
but these are generated from $f$-$F$ mixing from the allowed 
couplings $[FfH + FFH]_F$. It is intriguing that the cascade 
of (\ref{eq:EWSB}) for EWSB and (\ref{eq:FSB}) for FSB both require 
vector-like generations at the intermediate stage. We will argue, however,
that $F$ and $X$ cannot be identical. In particular, only $X$ has a large 
supersymmetry breaking mass, $[\bar{X} X]_A$, and only $F$ has direct
Yukawa couplings with ordinary matter $[\bar{F} \varphi f]_F$. These
distinctions, which arise because $F$ transforms non-trivially under $G_f$
while $X$ is trivial, result in an important difference between the last
cascade of (\ref{eq:EWSB}) and (\ref{eq:FSB}). In the case (\ref{eq:EWSB}),
the last cascade is induced by standard model gauge interactions, and leads to
positive squared masses for the scalar superpartners. On the other hand, the
last cascade of (\ref{eq:FSB}) is induced by the Yukawa couplings $[\bar{F}
\varphi f]_F$ and produces negative squared masses for $\varphi$, triggering
FSB. In our scheme, both mass scales of the EWSB sector, $M_{mess}$ and
$\langle H \rangle$, and both mass scales of the FSB sector, $M_F$ and
$\langle \varphi \rangle$, are generated via perturbative loops from the
single dynamical scale $\Lambda_{SSB}$, as illustrated in Figure 1.  We will
see later that $\langle \varphi \rangle / M_F \approx c/(16\pi^2)$, where $c$
represents a product of several coupling constants, so that $c$ may easily vary
between $1/10$ and $10$. This is sufficient for constructing viable models of
the fermion Yukawa matrices, like those in Refs.~\cite{LNS2,CHM2}.

It is well known that theories of weak scale supersymmetry also need mass
terms which couple the two Higgs doublets: $\mu [H_u H_d]_F$ and $m_3^2  [H_u
H_d]_A$. In supergravity theories, $\mu$ plausibly arises from higher order $D$
terms, giving $\mu = \Lambda_{SSB}^2 / M_{Pl}$, and a non-zero $\mu$ leads
automatically to a non zero $m_3^2$. With gauge mediated supersymmetry
breaking, $\Lambda_{SSB}$ is too small to allow such an origin for $\mu$: the
origin of $\mu$ and $m_3^2$ is problematic. Our scheme allows a simple origin
for both $\mu$ and $m_3^2$, as we will see in Section~3. 

The main accomplishment of this paper is the generation of flavor mass scales
$M_F$ and $\langle \varphi \rangle$ in a way which is analogous, but not 
identical, to the generation of $M_{mess}$ and $\langle H \rangle $. In 
Section 2 we survey the possible origins for fermion mass hierarchies in 
supersymmetric theories, and find that several alternative options are not 
very promising. In Section 3 we give the structure of the messenger sector 
at $M_{mess}$ and of the flavor sector at $M_F$. We discuss the differences 
between these sectors, showing that they cannot be identical. Keeping the 
group structure for $G_f$ general, we show explicitly how radiative 
corrections trigger the FSB vev $\langle \varphi \rangle$. We also study the 
EWSB sector in this framework.

The flavor physics scales of our framework are sufficiently low,
of order 10 TeV for $M_F$ and 100s of GeV for $\langle \varphi \rangle$, to be
both dangerous and interesting from the viewpoint of flavor changing and CP
violating processes. In Section 4 we study amplitudes for these processes
induced by integrating out the heavy vector generations, $F$, and the flavon
fields, $\varphi$.  These rare processes provide constraints on our framework
which are very different from the constraints they impose on supergravity
theories \cite{LNS2}. Hence the model building choices for the group $G_f$,
and for the representations of $\varphi, F$ and $f$, 
are governed by constraints which are summarized in the
conclusions, and which differ greatly from the supergravity case.

\section{Fermion Mass Hierarchy in Supersymmetric Models}
\setcounter{equation}{0}

The hierarchy in fermion Yukawa couplings is one of the major puzzles 
in the standard model.  The only known fermion with a Yukawa coupling
of order one is the top quark, while all other fermions have Yukawa
couplings that are significantly smaller\footnote{If $\tan\beta$
is large, the bottom quark and tau lepton may also have  
$O(1)$ Yukawa couplings.}.
The existence of small parameters in the standard model Lagrangian 
is natural in the sense of 't Hooft:  When all the Yukawa
couplings are set to zero, the standard model is invariant under a global
U(3)$^5$ flavor symmetry.  Thus, the symmetry of the theory is enhanced as
the Yukawa couplings are reduced.  While this explains why small Yukawa
couplings are technically natural, it sheds no light on how such 
small parameters arise. Thus, we would like to understand
how a low-energy effective theory containing small Yukawa couplings can 
arise when the corresponding high-energy theory involves no small parameters 
at all.

One way of framing this problem is to assume that the Yukawa 
couplings of the light fermions are forbidden at high energies by 
a flavor symmetry $G_f$, which is a subgroup of U(3)$^5$. The top quark Yukawa
coupling may be invariant under the flavor symmetry. Then, the problem at hand
is to understand why $G_f$ is broken only by a small amount.  Since the
flavor scale is generally much larger than the weak scale,
it is natural to work in the supersymmetric context.  Then the hierarchy
between the weak scale, the flavor scale, and any other high scales in the
problem will be stable against radiative corrections.

Supersymmetry, however, makes the task of generating small
couplings a challenging one. If small Yukawa couplings are
not present in the original superpotential, then the supersymmetric 
nonrenormalization theorem tells us that such couplings will {\em never} 
be generated at any order in perturbation theory.  Therefore, the
mechanism of flavor symmetry breaking must be linked either to
supersymmetric nonperturbative effects, or to supersymmetry breaking.  
Three popular schemes have been proposed for generating small 
Yukawa couplings in supersymmetric models: string compactification,
radiative generation, and the Froggatt-Nielsen mechanism.
Let us review these possibilities and consider the limitations 
of each:

{\em I. String Compactification}. It is possible that the small Yukawa
couplings are simply present as a boundary condition due to physics at the
string scale.  All coupling constants in string theory are supposed to be
proportional to a single string coupling constant, which is of the same order
as the gauge coupling constants, {\it i.e.}\/, ${\cal O}(1)$.  However,
couplings in the superpotential depend on the compactification.  In orbifold
models, if chiral fields belong to twisted sectors with different fixed
points, their superpotential couplings are suppressed by $e^{-R^2}$ in the
limit where the size of the compactified manifold $R$ is large \cite{Ibanez}. 
In this case, small numbers arise as a result of the compactification.  It 
has been pointed out that the radius $R$ does not need to be much larger
than the string scale \cite{Cassas}.  One possible problem with this scenario
is that different generations have different modular weights, and the scalar
masses are therefore non-universal. This may lead to dangerous 
flavor-changing neutral current effects \cite{Brignole2}. 

{\em II. Radiative Generation}.  If small Yukawa couplings are not already
present at the string scale, one may try to generate them
through radiative corrections that involve the soft supersymmetry-breaking 
operators. In scenarios of this type, the flavor 
symmetry $G_f$ is broken by the nonvanishing entries of the scalar mass 
matrices, while the fermion Yukawa matrices retain their flavor-symmetric form
at tree-level.  This can be a consequence of the spontaneous breakdown of the
flavor group. Yukawa couplings are then generated at the higher loop level
when the superpartners are integrated out.  The limitation of this
approach is that it is very difficult to generate small Yukawa couplings for 
both the first and second generation fermions, assuming the minimal particle
content of the supersymmetric standard model \cite{ACH1,ACH2}.  While the 
small first-generation Yukawa couplings may be understood as radiative 
effects, those of the second generation must be generated by a separate 
mechanism.  In  addition, the models that have been proposed require a 
special mechanism to ensure separate muon number conservation in the lepton
sector to evade the tight constraints from $\mu \rightarrow e\gamma$.  

{\em III. Froggatt-Nielsen Mechanism}. The Froggatt-Nielsen mechanism
\cite{FN} is perhaps the most popular mechanism for generating small
Yukawa couplings. Fields of the first and
second generations have no direct Yukawa couplings to the Higgs
bosons, as a consequence of a flavor symmetry. On the
other hand, heavy vector-like fields couple to the Higgs fields with
$O(1)$ strength.  When the flavor symmetry breaks spontaneously, a small
mixing is induced between the light generations and the vector-like fields, 
as described in the Introduction.  When the vector-like fields are
integrated out, small Yukawa couplings are generated in the low-energy
effective theory, having the form
\begin{equation}
\left(\frac{\langle \varphi \rangle}{M_F}\right)^n f f H \,\,\, .
\end{equation}
The smallness of the light fermion Yukawa couplings is a consequence of a
hierarchy between two scales $\langle \varphi \rangle / M_F$, where $M_F$ is
the mass scale of heavy vector-like fields, and $\langle \varphi \rangle$ is
the vacuum expectation value of a flavon field that spontaneosly breaks
$G_f$.  This is the mechanism of generating small Yukawa couplings 
that we will adopt in this paper.

There are two important questions associated with 
supersymmetric models that involve
the Froggatt-Nielsen mechanism. First, one may worry that the scalar mass
matrices may not be sufficiently degenerate to suppress flavor-changing
processes, since the three generations couple differently to the heavy states.
However, the flavor symmetry restricts the form of both the Yukawa and scalar
mass matrices, and this can be sufficient to prevent any flavor-changing
problems.  The flavor symmetry can either enforce a sufficient  degeneracy
among the scalar states, or align the Yukawa and scalar mass matrices so that
flavor changing neutral current processes are adequately suppressed
\cite{DLK,NS}.  In many models based on the Froggatt-Nielsen mechanism, one
typically needs $\langle \varphi \rangle / M_F \mbox{\raisebox{-1.0ex}
{$\stackrel{\textstyle ~<~}{\textstyle \sim}$}} 0.01$--0.05; powers of this 
ratio appear in the operators that generate the light fermion Yukawa
couplings. Thus, we are lead to the second question, which is more
fundamental: from where does the hierarchy $\langle \varphi \rangle / M_F \ll
1$ originate?  This is the main issue of the paper. 

Nonrenormalization theorems tell us that it is impossible to generate the
$\langle \varphi \rangle /M_F$ hierarchy perturbatively in models with
unbroken supersymmetry, unless the scales are put into the superpotential by
hand. This is exactly what we would like to avoid.  Therefore, the only
logical possibilities are that the origin of scales is either triggered by
({\em a}) non-perturbative effects or by ({\em b}) supersymmetry 
breaking (or both).  Let us consider each of these possibilities:

({\em a}) {\em Nonperturbative effects}.  One can imagine that the
Froggatt-Nielsen mass scale $M_F$ and the scale $\langle \varphi \rangle$
are generated by separate gauge groups that each become strong at
scales much lower than the Planck scale. The scale parameter of each
gauge group is given by $\Lambda_i = M_* e^{- 8 \pi^2 b_0^i / g_i^2}$, where
$g_i$ are the gauge coupling constants at $M_* = M_{Pl}/\sqrt{8\pi}$ and
$b_0^i$ are their beta function coefficients.  When the model is incorporated
into supergravity, a condensation due to strong gauge dynamics is likely to
break supersymmetry.  Even when the strongly interacting groups do not couple
directly to the fields in the MSSM, their scales have to be smaller than
$\Lambda_i \mbox{\raisebox{-1.0ex} {$\stackrel{\textstyle ~<~}{\textstyle
\sim}$}} (m_W M_{*}^2)^{1/3} \simeq 10^{13}$~GeV.  Thus, a small ratio in
scales, $\Lambda_1 / \Lambda_2 \sim 0.01$--0.05 requires a special
arrangement in the particle content and a mild fine-tuning in
the initial gauge couplings of both gauge groups.  The tuning becomes 
more and more severe as one considers lower scales for the $\Lambda_i$.

({\em b}) {\em Supersymmetry breaking.}  The other possibility is to use soft
supersymmetry breaking parameters to generate the scales $\langle \varphi
\rangle$ and $M_F$.  We discuss two cases separately.  The first possibility is
that supersymmetry breaking occurs in the hidden sector, and all soft
supersymmetry breaking terms are generated at the Planck scale, at the same
order of magnitude.  The other possibility is that supersymmetry breaking
occurs at low energy $\sim 10^7$~GeV, and is transmitted to the MSSM fields
via renormalizable interactions.  

The hidden sector case suffers from naturalness problems similar
to those that we encountered earlier. The basic difficulty is
that there is only one scale in the problem from which we would like to
generate two scales.  If all mass scales are generated by supersymmetry
breaking with a generic superpotential, they will all be of order
the weak scale, with no hierarchy among them.  Therefore, one needs to 
rely on flat directions of the superpotential to generate scales much 
higher than the weak scale. One may try to be economical by
identifying $M_F$ with $M_*$ (or $M_{\it string}$) and then hope to generate
$\langle \varphi \rangle \sim 10^{-2} M_*$.  A difficulty with this idea is that
one needs to forbid higher dimension operators in the superpotential of
the form $\varphi^{n+3}/M_*^n$, up to $n \mbox{\raisebox{-1.0ex}
{$\stackrel{\textstyle ~>~}{\textstyle \sim}$}} 7$ in order that 
$\langle \varphi \rangle$ not be pushed down to lower scales.  If the flavor
symmetry is $Z_N$, then one needs a relatively large $N
\mbox{\raisebox{-1.0ex} {$\stackrel{\textstyle ~>~}{\textstyle \sim}$}} 10$. 
If the flavor symmetry is continuous, it needs to be gauged or it will be
violated by Planck scale effects.  If it is a gauged U(1) symmetry, we always 
require fields with positive and negative charges to cancel the anomaly.
As a result, it is likely that there will be higher dimension operators
allowed by the gauge symmetry that reduce $\langle \varphi \rangle$.  This
statement trivially extends to non-Abelian gauge symmetries as well. 
Therefore, a gauged flavor symmetry had better be anomalous, with the anomaly
cancelled by the Green--Schwarz mechanism.  This possibility has been studied
by many authors, who have focused on obtaining the correct $\sin^2
\theta_W$=3/8  from the anomaly cancellation condition \cite{many}.  In general,
however, this scenario leads to non-degenerate scalar masses because the U(1)
charge assignments are generation-dependent. \cite{DPS2}.  If there is quark-
squark alignment \cite{NS}, then scalar nondegeneracy may not lead to
dangerous flavor changing effects. However, we are not aware of any models in
which alignment is achieved using the same U(1) whose anomalies are cancelled
by the Green--Schwartz mechanism.  It remains to be seen whether a model of
this type can be constructed.

The obvious way to avoid the problems with higher dimension operators is
to lower the scales of $M_F$ and $\langle \varphi \rangle$.  The potential
along a flat direction is given by $m^2(\varphi) |\varphi|^2$ where $m^2
(\varphi)$ is the effective supersymmetry breaking squared mass which
satisfies the renormalization group equation.  The minimum of the
potential is generated around the scale where $m^2 (\varphi)$ crosses zero.
In this way, one can easily obtain the invariant mass $M_F$ of the vector-like
Froggatt-Nielsen fields such that $M_F \ll M_{Pl}$; then higher 
dimension operators become completely irrelevant.  Once the scale
$M_F$ is generated, the vector-like fields decouple from the renormalization
group equation of the flavon mass squared.  Therefore the running of the
flavon mass squared is slowed, and it crosses zero at a much lower
scale, which is likely to be much less than one hundreth of $M_F$. 
To obtain $\langle \varphi \rangle / M_F \sim 0.01$ requires the model
to be very carefully arranged.

Models with low-energy supersymmetry breaking do not suffer from the 
difficulties discussed above.  While supersymmetry breaking in the 
hidden sector presented us with only one scale from which
we needed to generate two, low-energy supersymmetry breaking mediated 
by renormalizable interactions tends to produce a multitude of scales.  
Since supersymmetry breaking is mediated to the MSSM fields, or to any 
other fields in the theory, via renormalizable interactions, 
the effects are not always transmitted at the same order in perturbation
theory. Thus, it is natural to obtain many different mass scales
separated from each other by powers of $1/(16\pi^2)$.

This observation suggests an intriguing scenario: Supersymmetry is 
broken at a scale $\Lambda_{SSB} \sim 10^7$~GeV.  This scale is
determined by dimensional transmutation, and is not directly
input into the theory.   Supersymmetry breaking is mediated via
renormalizable interactions to all other fields in the theory.  
This occurs at varying order in perturbation theory, producing
a hierarchy of scales, of the form $(1/16 \pi^2)^n \Lambda_{SSB}^2$.  Thus, 
the small flavor symmetry breaking parameters described earlier are 
identified with the ratio of some of these scales.  Notice
that the range of $\langle \varphi \rangle / M_F$ that is required
by the Froggatt--Nielsen mechanism, $\simeq 0.01$--0.05 
corresponds to a loop factor ($1/4\pi$ or $1/16\pi^2$) times an order 
one coefficient.   In this framework, all coupling constants in the 
superpotential can be ${\cal O}(1)$, and all mass scales generated
from a single scale, the scale at which supersymmetry is broken.  
 
In the rest of the paper we show how our framework may be implemented.  We
focus on the generic structure of our framework and demonstrate that it is 
phenomenologically viable.  We do not go into a detailed discussion of particular
flavor symmetries.  Our framework is compatible with a variety of explicit
flavor models that are described in the literature, including $U(1)^3$
\cite{NS,LNS2}, $\Delta(75)$ \cite{KS}, and $(S_3)^3$ \cite{HM,CHM2,CHM1}. 

\section{Framework} \label{sec:frame}
\setcounter{equation}{0}

The overall structure of our model is summarized schematically in Figure~1.
Supersymmetry is broken at the scale $\Lambda_{SSB}$, and communicated
to both the gauge-mediation (GM) and flavor sectors via 
two-loop diagrams.  We assume that different U(1) gauge interactions act 
as the messengers of supersymmetry breaking to each of these sectors.
Thus, the GM and flavor mass scales are given by $\Lambda_{GM} \approx 
g_m^{\prime 2} \Lambda_{SSB}/16\pi^2$ and $\Lambda_{\it flav} \approx 
g_m^2 \Lambda_{SSB}/16\pi^2$, where $g'_m$ and $g_m$ are the
messenger U(1) gauge couplings\footnote{$\Lambda_{GM}$ and $\Lambda_{\it flav}$
also depend on couplings in the supersymmetry breaking sector which we 
have omitted for simplicity.}.  The ordinary superparticles $\tilde{f}$ 
and the flavor symmetry breaking fields $\varphi$ communicate with
the supersymmetry breaking sector through four-loop diagrams, and develop
masses of order $g_{m}^{\prime 2} \Lambda_{SSB}/(16\pi^2)^2$ and
$g_{m}^2\Lambda_{SSB}/(16\pi^2)^2$, respectively.  The ordinary
Higgs field $H$ develop masses comparable to those of $\tilde{f}$ for the
choice $g'_m > g_m$ that we assume below.

\subsection{Constraints on the Flavor Sector} \label{subsec:con}

The requirement that we generate TeV scale masses for the ordinary
superparticles fixes $\Lambda_{GM}$ at approximately $100$ TeV.  
This is consistent with a supersymmetry breaking scale $\Lambda_{SSB}\approx
10,000$ TeV.  On the other hand, we choose the flavor scale to be
somewhat lower, $\Lambda_{\it flav}\approx 10$ TeV, so that we do not
generate negative squared masses for the Higgs fields that are too
large (see Section 3.5).  In addition, this choice reduces flavor changing
neutral current effects that originate from the supersymmetry-breaking masses
of the Froggatt-Nielsen (FN) fields (see Section 3.7). The
difference between the GM and flavor scales can be obtained by choosing the
messenger U(1) gauge couplings such that $g_m/g'_m \approx 1/3$, 
which does not constitute a significant fine-tuning\footnote{While we have
introduced two messenger U(1) gauge groups to obtain different flavor and
GM mass scales, the gauge structure of our model has an additional benefit. 
If there were only one U(1) gauge interaction coupling to the two otherwise
disconnected sectors of the model, we would be left with an extra
Nambu-Goldstone boson after symmetry breakdown.  While it is 
not clear whether such a massless scalar boson would do any harm 
phenomenologically, we have chosen to avoid this situation completely: the
Nambu-Goldstone boson is absorbed by the additional U(1) gauge field.}.  With
the flavor scale at $10$ TeV, the flavon fields $\varphi$ develop masses of
order a few hundred GeV. We will show in Section~4 that the possibility of
flavon fields with masses in the few hundred GeV range is not excluded by the
current phenomenological constraints.

In order to generate the flavor scale in the way suggested above, we must 
first address some immediate phenomenological difficulties.
Let us assume that there is a field $a$ whose vacuum expectation value (vev)
generates the mass of the FN fields.  Consider the superpotential couplings
\begin{equation}
W = \alpha a F \overline{F} + \beta \overline{F} \varphi f \,\,\, ,
\label{eq:start}
\end{equation}
where $F$ and $\overline{F}$ are FN fields, $\varphi$ is a flavon,
and $f$ is an ordinary matter field.  The first term determines
the FN mass scale $M_F = \alpha \langle a \rangle$, while
the second term generates the desired mixing between the ordinary fields
and the heavy FN fields beneath the flavor-symmetry-breaking scale.  Since 
we assume that the vev of $a$ is a consequence of supersymmetry breaking, 
we expect that the auxiliary component of $a$ will also be 
nonvanishing, and in general,
\begin{equation}
\langle F_a \rangle \approx \langle a \rangle^2 \,\,\, .
\label{eq:bad}
\end{equation}
Now consider the scalar mass squared matrix for the ordinary and
the FN fields.  If $a$ has a nonvanishing $F$ component, then there will
be a scalar $B$-term of the form $F\overline{F}$ of order 
$\alpha \langle F_a \rangle \approx M_F^2$. From Eq.~(\ref{eq:start}), the 
scalar mass squared matrix is then given by
\begin{equation}
\left(\begin{array}{ccc}  M_F^2 & \alpha \langle F_a \rangle & 0  \\
\alpha \langle F_a \rangle  & M_F^2 & \beta M_F \langle \varphi \rangle \\
0     &\beta M_F \langle \varphi \rangle &  \beta^2
\langle \varphi \rangle^2 \end{array} \right)
\label{eq:negeig}
\end{equation}
in the basis $(\overline{F}^*, F, f)$.  In the case where 
$\langle F_a\rangle = 0$, the matrix (\ref{eq:negeig}) has one
zero eigenvalue (corresponding to the physical squarks or sleptons), and 
two eigenvalues of order $M_F^2$.  When $F_a$ is nonvanishing, the zero 
eigenvalue is shifted to
\begin{equation} 
-\beta^2 \left(\frac{\langle \varphi \rangle}{M_F}\right)^2 
\langle F_a \rangle ^2/M_F^2 \,\,\,. 
\label{eq:prob}
\end{equation}
The fact that the lightest eigenvalue is negative is not a problem by 
itself, since there are larger positive contributions to the squared 
masses from the gauge-mediation diagrams in the gauge mediation
sector of the model, as we will see later.  However, the contribution in
Eq.~(\ref{eq:prob}) is {\em flavor dependent}, and can lead to large flavor
changing neutral current effects. The simplest way of avoiding these
difficulties is to construct models in which $F_a$ is naturally much smaller
that $\langle a \rangle ^2$, so that the effect of Eq.~(\ref{eq:prob}) is
phenomenologically irrelevant\footnote{If the squarks have masses 
around $1$ TeV, then flavor-changing neutral current (FCNC) effects in
the quark sector due to 
Eq.~(\ref{eq:prob}) may not necessarily be fatal.  For example, in an 
explicit model of flavor with $\langle \varphi \rangle /M_F \sim 1 \times
10^{-2}$, the (1,2) elements of the squark mass squared matrices will be of
the order $\tilde{m}^2_{12}/\tilde{m}^2 \approx 0.01 $, assuming that
$F_a \approx \langle a \rangle^2 =$ ($10$ TeV)$^2$.  This is in borderline 
agreement with the current experimental bounds.  The real problem arises
is the lepton sector, where the right-handed sleptons are a factor
of $7$ lighter than the squarks.   Lepton flavor violation will be present 
at an unacceptable level unless a separate FN sector is constructed for the
leptons that preserves electron or muon number.  The solution presented in 
the text does not place additional restrictions on the flavor structure
of the model.}.
  
Thus, we choose to build a model in which the messenger sector 
for generating the FN mass scale has some mechanism of protecting 
the $a$ field from acquiring an $F$ component, at least at tree level.  
The situation is quite different in the gauge mediation sector, where the
field analogous to $a$ must acquire both scalar and $F$-component vevs to
produce the correct nonsupersymmetric spectrum of vector-like states
\cite{DN}.  Thus, in addition to a differing sequence of mass scales, the two
branches of Figure~1 are distinguished by the properties of the field that
couples to the vector-like multiplets; thus, the GM and flavor sectors cannot
be identified.  We will see this explicitly in the model that we present
below. 

\subsection{The Supersymmetry Breaking Sector} \label{subsec:dsb}

We assume that supersymmetry is broken in a sector of the model that is nearly
isolated from all other sectors.  The only communication between fields 
in the supersymmetry breaking sector and the remaining fields in the 
theory is through the two messenger U(1) gauge interactions.  Fields 
$\xi_\pm$ that carry either of the messenger charges can communicate with 
the supersymmetry breaking sector through two-loop diagrams.  When 
supersymmetry is broken, the $\xi_\pm$ fields can acquire supersymmetry
breaking masses.  In some models of dynamical supersymmetry breaking, 
like the SU(6)$\times$U(1) model discussed in Refs.~\cite{DN,DNNS}, 
it is known that the $\xi$ fields can acquire negative squared masses 
after supersymmetry is broken.  Although our framework involves 
two messenger U(1) gauge groups, we assume that the same is possible here.
We view this as a mild restriction on the types of model that can serve as 
an adequate supersymmetry breaking sector. Note that there are many 
models of dynamical supersymmetry breaking that contain two 
nonanomalous U(1) factors that can be gauged.  Examples include 
the SU(9) model with an anti-symmetric tensor and five 
anti-fundamentals \cite{ADS2}, and the SU(2) model with four doublets 
and six singlets \cite{IZYAN}.  Both of these models possess large 
global symmetries (SP(4) or SU(4)) that contain a nonanomalous 
U(1)$\times$U(1) subgroup \footnote{In addition, Fayet--Illiopoulos $D$-terms 
are not generated in these models because of unbroken discrete 
symmetries \cite{DNNS}.}.  It is reasonable to assume that in some of 
these models, there are regions of parameter space in which it is
possible to generate negative squared masses for fields carrying 
either of the messenger U(1) charges.  This point will be assumed in
the next two subsections.

\subsection{Gauge Mediation Sector} \label{subsec:gauge}

We first consider the sector of the theory that generates the
gauge mediation mass scale, following the work of Dine, Nelson, Nir and
Shirman \cite{DNNS}. The superpotential for this sector is given by 
\begin{equation}
W = - h'_1 \xi'_+\xi'_- S'_1 + \frac{\lambda'_1}{3} S_1^{\prime 3} + 
\alpha'_1 S'_1 \overline{X} X 
\label{eq:ndsup}
\end{equation}
where $X$ and $\overline{X}$ are vector-like fields that carry standard
model quantum numbers, and $S'_1$ is a gauge singlet.  The $\xi'$ fields
are charged under the messenger U(1) gauge group that connects the
superpotential above to the supersymmetry breaking sector of the theory.
To prevent the fields in Eq.~(\ref{eq:ndsup}) from coupling to fields in
the flavor sector, we will impose a $Z_3$ ``sector symmetry'' under
which all the GM sector fields transform by the phase
$e^{2i\pi/3}$.\footnote{Without this symmetry, the singlet $S'_1$ above
could also couple to the FN fields.  Then we would not be able to avoid
the phenomenological disasters described in Section~\ref{subsec:con}.}
Then (\ref{eq:ndsup}) includes the most general renormalizable
interactions consistent with the symmetries of the theory.  The two-loop
diagrams described in Section~\ref{subsec:dsb} contribute to the
$\xi'_+$ and $\xi'_-$ masses after supersymmetry is broken. Thus, in
addition to the superpotential given above, we assume there are soft
supersymmetry-breaking masses
\begin{equation}
V = - {m'}^2 |\xi'_+|^2 - {m'}^2 |\xi'_-|^2  \,\,\, .
\end{equation} 
With $\Lambda_{SSB} \approx 10,000$ TeV, we expect the $\xi'_\pm$ masses $m'$
to be of order  $(\alpha'_m/4\pi) \Lambda_{SSB} \approx 100$ TeV, where 
$\sqrt{4\pi \alpha} = g'_m$ is the relevant messenger U(1) gauge coupling.  
The negative squared masses for the $\xi'_\pm$ generate nonvanishing vevs 
for both the scalar and $F$ components of $S'_1$, of order $m'$ and
$m^{\prime 2}$
respectively.  This leads to a nonsupersymmetric spectrum for the 
fields $X$ and $\overline{X}$, which can communicate with the ordinary
fields $\tilde{f}$ via standard model gauge interactions.

If the $X$ and $\overline{X}$ form a {\bf 5} and ${\bf \overline{5}}$ of 
SU(5), then the ordinary gaugino and squark masses are given by \cite{DN}
\begin{equation}
m_{i}=\frac{g_i^2}{16 \pi^2} \frac{\langle F_{S'_1} \rangle}
{\langle S'_1 \rangle} \,\, ,
\label{eq:mgaugino}
\end{equation}
\begin{equation}
\tilde{m}^2 = \sum_i 2 C_F^{(i)} \left( \frac{g_i^2}
{16 \pi^2}\right)^2 \frac{\langle F_{S'_1} \rangle^2}
{\langle S'_1 \rangle^2} \,\, ,
\label{eq:msquark}
\end{equation}
where the $g_i$ are standard model gauge couplings, and the Casimir 
$C_F$ is 3/4 for SU(2) doublets, 4/3 for SU(3) triplets, and (3/5)$Y^2$ 
for fields with ordinary hypercharge $Y$.  For our choice
$\Lambda_{GM} = \langle F_{S'_1} \rangle/\langle S'_1 \rangle = 100$~TeV,
we obtain the the squark and slepton masses (at 100 TeV),
\begin{displaymath}
\begin{array}{ccccc}
\tilde{q} & \tilde{u} & \tilde{d} & \tilde{l} & \tilde{e}\\ \hline
1140 & 1100 & 1100 & 350 & 150
\end{array} \,\,\,
\end{displaymath}
and the gaugino masses (at 1 TeV)
\begin{displaymath}
\begin{array}{cccc}
\tilde{B} & \tilde{W} & \tilde{g} \\ \hline
130 & 270 & 930
\end{array} \,\,\, ,
\end{displaymath}
in GeV. Notice that gauge-mediation renders the squarks and gluinos 
heaviest, with masses around $1$ TeV.

It is important to point out that there are a number of problems with 
the messenger superpotential in Eq.~(\ref{eq:ndsup}), that have not 
been discussed in the literature.  First, the coupling $h'_1$ must be
taken to be 
smaller than $g'_m$ so that the $D$-term vanishes at the
minimum of the above potential.  Second, the coupling $h'_1$ must be 
also smaller than $\lambda'_1$ so that the desired vacuum with $\langle
S \rangle \neq 0$ is a 
local minimum.  This is in disagreement with the claim made in 
Ref.~\cite{DNNS}.  The region $\alpha'_1 \lambda'_1/(\lambda'_1+\alpha'_1) 
< h'_1 < \lambda'_1$ is also problematic, because $\alpha'_1 \langle F_{S'_1}
\rangle$  will be larger than $(\alpha'_1 \langle S'_1 \rangle)^2$, and the
scalar mass squared matrix for the $X$ and $\overline{X}$ fields will have a 
negative eigenvalue.  This would imply that the standard model gauge 
group is broken at the $100$ TeV scale.  Therefore, we need $h'_1 < 
\alpha'_1 \lambda'_1/(\lambda'_1+\alpha'_1)$.  Even in this region, 
there is still a global minimum of the potential where the standard 
model gauge group is broken.  To see this, notice that in the 
supersymmetric limit there is a flat direction in which $S'_1 = 0$, 
$\xi'_+ = \xi'_-$, $X = \overline{X}$, and $h'_1 \xi'_+ \xi'_- = \alpha'_1 
\overline{X} X$.  Along this direction, the potential becomes 
increasingly negative until the negative mass squared for the $\xi'$ 
fields $- m^{\prime 2}$ disappears beyond $\xi'_\pm \approx \Lambda_{SSB}$.  
For larger values of $\xi'$, the potential becomes flat with $V \simeq 
- m^{\prime 2} \Lambda_{SSB}^2$, which is much lower than the local minimum, 
where $V \simeq - m^{\prime 4}$.  In order to avoid this problem, it seems 
that the gauge mediation sector must be modified.  One obvious 
solution is to introduce an another singlet field $S'_2$ with the the 
additional superpotential couplings
\begin{equation}
\Delta W = - h'_2 \xi'_+\xi'_- S'_2 + \frac{\lambda'_2}{3} S_2^{\prime 3} + 
\alpha'_2 S'_2 \overline{X} X + \mbox{($S'_1$, $S'_2$ couplings)} .
\end{equation}
Unless $h'_2/\alpha'_2 = h'_1/\alpha'_1$, there is no longer a flat
direction.  The desired vacuum can be a global minimum in a certain
range of parameters as discussed above.  The precise conditions on the
parameters are complicated and not worth  presenting here.  It
should be stressed that the details of the gauge mediation sector are
irrelevant to the flavor sector presented in the next subsection.  Thus,
any workable alternative to Eq.~(\ref{eq:ndsup}) can be adopted without
altering the conclusions of this paper.

\subsection{The Flavor Sector} \label{subsec:flav}

As we described in Section~\ref{subsec:con}, the field that determines 
the FN mass scale, $a$, must have no $F$-component at tree level.  
If a self-coupling term $\lambda a^3$ were present in the superpotential,
then we would expect the $a$ field to acquire an $F$ component
$F_a = 3 \lambda a^2 \neq 0$ when $a$ acquires a vev.  Therefore,
we will begin by imposing a $Z_2$ symmetry under which the $a$ field is 
odd. In the superpotential that we present below, this will be sufficient
to prevent $F_a$ from acquiring a vev at lowest order in perturbation
theory.

Since we would like to generate $\langle a \rangle$ from supersymmetry
breaking, the field $a$ must couple, at least indirectly, to fields that
are charged under the messenger U(1).  The simplest renormalizable
superpotential that generates a vev for $a$ is
\begin{equation}
W = h_1 \xi_+ \xi_- S_1 - g_1 S_1 a^2 + 
\frac{\lambda_1}{3} S_1^3 
\,\,\, ,
\label{eq:Wxi}
\end{equation}
where the $\xi_\pm$ and $S_1$ fields are even under the $Z_2$,
and we allow no dimensionful couplings.  As before, we assume
that the $\xi$ fields develop negative squared masses when 
supersymmetry is broken,
\begin{equation}
V = - m^2 |\xi_+|^2 - m^2 |\xi_-|^2  \,\,\, .
\end{equation}
where $m$ is of order $(\alpha_m/4\pi)\Lambda_{SSB} \approx 10$ TeV.
Thus, the $\xi_\pm$ fields acquire vevs $\langle \xi_+ \rangle 
= \langle \xi_- \rangle = \langle \xi \rangle$ which are of order
$m$.  Given these vevs, the $S_1$ field develops a mass 
$h_1 \langle \xi \rangle$.  As long as $\lambda_1 
< h_1$, $S_1$ remains at the origin, while  
$F_{S_1} = h_1 \langle \xi \rangle^2 \neq0$.  Since 
$F_{S_1} \neq 0$, the scalar $(a,a^*)$ mass squared matrix
develops a negative eigenvalue, and $a$ then obtains a vev.  
However, $S_1$ does not have a vev in this parameter range, 
so $F_a$ exactly vanishes, as desired.

Unfortunately, this simple superpotential has the same problem that
we encountered with the first superpotential presented in 
Section~\ref{subsec:gauge}: there is a flat direction where 
$h_1\xi_+\xi_- = g_1 a^2$.  As a consequence of the negative
squared masses of the $\xi_\pm$ fields, the potential has a running-away
behavior along the flat direction, and thus we expect that 
$\langle \xi \rangle$ will be at least of order $\Lambda_{SSB}$. To
eliminate the unwanted flat direction, we introduce a new 
field, $S_2$, with the following superpotential interactions:
\begin{equation}
\Delta W = h_2 \xi_+ \xi_- S_2 - g_2 S_2 a^2 +
\frac{\lambda_2}{3} S_2^{3} +  \mbox{($S_1$, $S_2$ couplings)}
\,\,\, .
\label{eq:DWxi}
\end{equation}
Again, all the fields above are even under the $Z_2$ symmetry, except
for $a$.  The coupling $\alpha a \overline{F} F$ generates the desired 
masses of the FN fields, without the bilinear supersymmetry breaking
mass terms.  Notice that the superpotential in the flavor sector
is actually identical to the one in the gauge mediation sector providing
we make the identification $\overline{X} X \leftrightarrow a^2$.  The
only difference between these sectors is the allowed range of the
superpotential couplings.  In the flavor sector we take the
$\lambda_i$ to be smaller than the $h_i$, so that $S_1$ and $S_2$ 
will not develop vevs.  In the gauge-mediation sector, we take the 
$\lambda'_i$ to be larger than the $h'_i$ so that $S'_1$ and $S'_2$ 
fields do acquire vevs, while $X$ and $\overline{X}$ do not.

To study the scalar potential of the theory, it is convenient to
work with the redefined superpotential couplings
\begin{eqnarray}
W &=& h \xi_+ \xi_- \chi + g \xi_+ \xi_- \eta - \lambda a^2 \eta +
\nonumber \\
& & + \frac{k_1}{3} \chi^3 + k_2 \chi^2 \eta + k_3 \chi
\eta^2 
+ \frac{k_4}{3} \eta^3 \,\,\, ,
\label{eq:fnsup}
\end{eqnarray}
where $\chi$ and $\eta$ are linear combinations of $S_1$ and $S_2$.
$h$, $g$ and $k_i$ ($i = 1,2,3,4$) are coupling constants of the
redefined fields.  Using suitable phase rotations of fields, we can take
$h$, $g$ and $\lambda$ real and positive without a loss of generality.
The scalar potential of the theory is given by
\begin{eqnarray}
V &=& |h\xi_+\xi_- + k_1 \chi^2 + 2 k_2 \chi \eta + k_3 
\eta^2|^2 + |\xi_+(h\chi+g\eta)|^2 + |\xi_-(h\chi+g\eta)|^2
\nonumber \\
& & +|g\xi_+\xi_--\lambda a^2 + k_2 \chi^2 + 2 k_3 \chi \eta
+k_4 \eta^2|^2 + |2\lambda a \eta|^2
\nonumber \\
& & + \frac{g_m^2}{4}(|\xi_+|^2-|\xi_-|^2)^2 - m^2 |\xi_+|^2 - m^2 |\xi_-|^2
\,\,\, .
\label{eq:FNpotential}
\end{eqnarray}
For simplicity, we assume $g_{m} > h$.  Then the potential is 
minimized when the messenger U(1) $D$-term contribution to the 
potential is minimized: $\langle \xi_+ \rangle = \langle \xi_- 
\rangle$.\footnote{If $g_{m} < h$, the potential prefers $\langle 
\xi_{+} \rangle \neq \langle \xi_{-} \rangle$ and develops a different 
minimum.} In some region of parameter space,
the rest of the potential is minimized when 
$\langle \eta \rangle = \langle \chi \rangle =0$, and 
\begin{equation}
\langle a \rangle = \frac{1}{h}\sqrt{\frac{g}{\lambda}} \, m
\,\,\,\,\,\,\mbox{ and } \,\,\,\,\,\,
\langle \xi_\pm \rangle = \frac{1}{h} \, m \,\,\, . 
\end{equation}
It is straightforward to check that the $F$ components of $a$, $\xi_+$, 
$\xi_-$ and $\eta$ vanish at this minimum; $\chi$ develops an $F$ 
component $F_\chi = m^2/h$.  Of course, there are corrections to
these results that are suppressed by loop factors. For example,
we expect supersymmetry breaking to generate soft trilinear terms in addition
to the squared masses of the $\xi_\pm$ fields.  Since these terms appear
at the same order in the loop expansion as the $\xi_\pm$ masses,
they will have coefficients $A\approx\Lambda_{SSB}/(16\pi^2)^2$, which
implies that $A/m \approx 1/(16\pi^2)$.  In this case, we find
that a nonvanishing $F$ component of $a$ is indeed generated. However, 
it is more than adequately suppressed: 
$\langle F_a \rangle \approx \langle a \rangle^2/(16\pi^2)$.

Given these vevs, we can now study the effects of supersymmetry breaking
on the scalar and fermion spectra of our model.  Only the fields 
that mix with $a$ are relevant to the the generation of flavon masses, as
we will see below. It is straightforward to show that the scalar and fermion 
masses matrices are two-by-two block diagonal in the basis 
($\xi$,$a$,$\chi$,$\eta$,$\xi'$), where $\xi = (\xi_++\xi_-)/\sqrt{2}$
and $\xi' = (\xi_+-\xi_-)/\sqrt{2}$.  Since $\xi$ is the only field
that mixes with $a$, we focus on the $(\xi, a)$ submatrices. In this basis, 
the mass squared matrices for the real and imaginary scalar components are 
given by
\begin{equation}
M^2_r = \left[\begin{array}{cc}
2h^2+2 g^2 & -2 g\sqrt{2\lambda g} \\
-2 g\sqrt{2 \lambda g} & 4 g\lambda 
\end{array}\right]\frac{m^2}{h^2}  
\label{eq:mr}
\end{equation}
and 
\begin{equation}
M^2_i = \left[\begin{array}{cc}
2 g^2 & -2 g\sqrt{2\lambda g} \\
-2 g\sqrt{2 \lambda g} & 4 g\lambda \\
\end{array}\right]\frac{m^2}{h^2} \,\,\,\, ,
\label{eq:mi}
\end{equation}
while the squared mass matrix for the fermionic components is
\begin{equation}
M^2_f = \left[\begin{array}{cc}
2 h^2+2 g^2 & -2 g\sqrt{2\lambda g} \\
-2 g\sqrt{2 \lambda g} & 4 g\lambda \\
\end{array}\right]\frac{m^2}{h^2} \,\,\, .
\label{eq:mf}
\end{equation}   
Notice that $M^2_r$ and $M^2_f$ are identical, while $M^2_i$ differs in its
$(1,1)$ component.  As far as we are interested in  the effects of
supersymmetry breaking in the $\xi$-$a$ spectrum alone, we may evaluate the
deviation of a given Feynman diagram about its supersymmetric limit by
replacing the propagator for the imaginary components of $\xi$ and $a$ by the
difference 
\begin{equation}
i (p^2 - M^2_i)^{-1} - i (p^2 - M^2_f)^{-1} \,\,\, . 
\label{eq:propdiff}
\end{equation}

The potential (\ref{eq:FNpotential}), combined with a term $\alpha 
a \overline{F} F$, generates the supersymmetric mass 
of FN fields while avoiding supersymmetry breaking bilinear term.
This is a consequence of the $Z_{2}$ symmetry that we imposed at the start. 
For simplicity, we assume that the FN fields $F$ and ordinary matter fields
$f$ transform in the same way under this $Z_2$.  The couplings
$a \overline{F} F$ and $\overline{F} \varphi f$ then imply that the
flavons $\varphi$ are necessarily odd under this $Z_{2}$.  

\begin{figure}[t]
	\centerline{\psfig{file=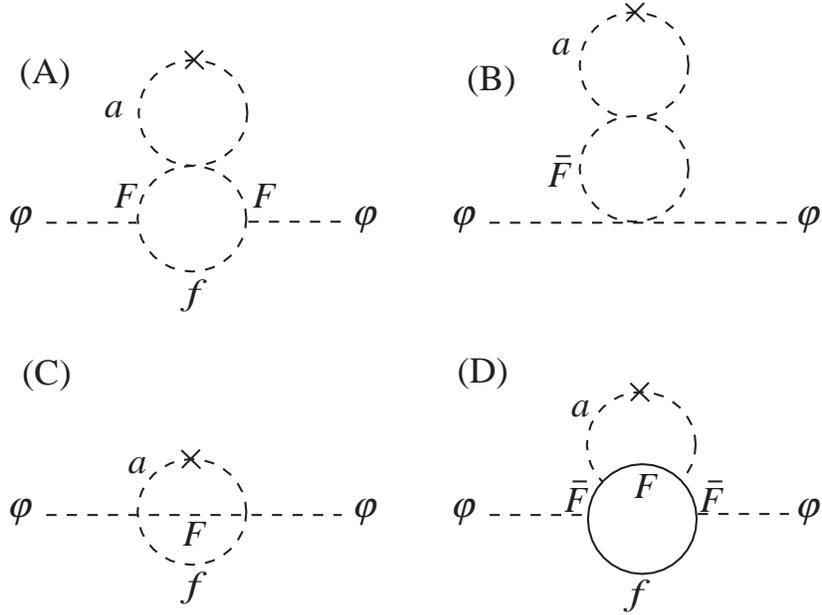,width=0.8\textwidth}}
	\caption[1]{Two-loop Feynman diagrams that generate the negative
	squared flavon masses.  The cross on the $a$ propagator indicates
	the supersymmetry breaking effect given in (\ref{eq:propdiff}).
	See text for more details.}  
	\protect\label{flavon}
\end{figure}

Now we are in a position to show that supersymmetry breaking
generates a negative mass squared for the flavon fields
$\varphi$.  To communicate flavor symmetry breaking to the
ordinary fields, we assume the superpotential couplings
\begin{equation}
W_\varphi = \beta \overline{F} \varphi f + \gamma \varphi^2 \varphi' 
+ \delta \varphi^{\prime 3}/3 \,\,\,.
\label{eq:wphi}
\end{equation}
where $\beta$, $\gamma$, and $\delta$ are coupling constants.  The
fields $\varphi'$ are even under the $Z_2$, and hence do not
have an $\overline{F} \varphi' f$ coupling.  The interactions in
Eq.~(\ref{eq:fnsup}) and Eq.~(\ref{eq:wphi}) allow us to write down the
two-loop diagrams presented in Figure~2.  Only the diagrams that involve the
imaginary part of $a$ are shown.  The cross on the $a$ line indicates we are
using the $(2,2)$ element of the difference propagator in
Eq.~(\ref{eq:propdiff}).  The details of this calculation are 
presented in the Appendix. The result is
\begin{equation}
m^2_\varphi \approx - \frac{N_c}{(16\pi^2)^2} \,
\frac{\alpha^2 \beta^2 g^3 \lambda}{h^4}
\, m^2  \,\,\, ,
\label{eq:result}
\end{equation}
where $N_c$ is the number of colors, and we have assumed that all 
superpotential couplings are of order unity. This generates the vacuum 
expectation values 
\begin{equation}
      \langle \varphi \rangle^{2} = 
      \frac{\delta^{2}}{8 \gamma^{3} (\delta-\gamma)} (- m_{\varphi}^{2})\, ,
      \hspace{1cm}
      \langle \varphi' \rangle^{2} =
      \frac{\delta - 2\gamma}{8 \gamma^{2} (\delta - \gamma)} (-
m_{\varphi}^{2})
      \, ,
      \label{eq:ep1}
\end{equation}
if $\delta > 2 \gamma$, or
\begin{equation}
      \langle \varphi \rangle^{2} = 
      \frac{1}{2 \gamma^{2}} (- m_{\varphi}^{2})\, ,
      \hspace{1cm}
      \langle \varphi' \rangle^{2} = 0
      \, ,
      \label{eq:ep2}
\end{equation}
if $\delta < 2 \gamma$.  For the first choice of parameters, $\varphi'$
acquires a vev.   We will see in Section 4.1 that this may be
desirable in some cases to avoid  model-dependent pseudo-Nambu-Goldstone
bosons that are too light.  
With the second choice of parameters, $\varphi'$ does not acquire a
vev, and $F_\varphi$ remains vanishing. This choice will be useful 
when we need to suppress the flavor-dependent trilinear soft supersymmetry
breaking terms generated when $F_\varphi \neq 0$; these will be discussed in
Section~\ref{subsec:fcp}.  The exact pattern of $\varphi'$ vevs and 
$F$-components is a highly model-dependent issue.  Eqs.~(\ref{eq:ep1})  
and (\ref{eq:ep2}) allow us to leave our options open;  it is quite 
likely that the relative size of $\gamma$ and $\delta$ differs from 
one flavon to the other. 

Eqs.~(\ref{eq:ep1},\ref{eq:ep2}) are the origin of the hierarchical 
pattern of flavor symmetry breaking in our model.  Notice that 
Eqs.~(\ref{eq:result},\ref{eq:ep1},\ref{eq:ep2}) depend on the product 
of many different coupling constants, so that the precise ratio 
$\langle\varphi\rangle/m$ can vary significantly from one flavon to 
the other, even when the individual couplings are not far from unity.  
Thus, it is possible that the flavons coupling to leptons may have 
vacuum expectation values that are systematically smaller than those 
coupling to quarks, as a consequence of the different renormalization 
group running of the corresponding superpotential couplings.  This may 
partially alleviate the more severe flavor-changing problem in the 
lepton sector.  The typical parameter range we have in mind is 
$m_{\varphi}^{2} \sim (400~\mbox{GeV})^{2}$ in the quark sector and 
$\sim (100~\mbox{GeV})^{2}$ in the lepton sector; with 
the coupling constants $\gamma, \delta$ varying between 1/3 to unity, we 
obtain $\langle \varphi \rangle \sim 0.4$--$1$~TeV 
and $\langle \varphi \rangle \sim 100$--$300$~GeV, respectively.
  
To properly implement the FN mechanism, we must also include the Yukawa 
couplings of the ordinary Higgs fields $H$
\begin{equation}
W_H = \zeta F H f + \kappa F F H + h_t f f H \,\,\, ,
\label{eq:whiggs}
\end{equation}
where $\zeta$ and $\kappa$ are coupling constants.  The term 
proportional to $h_t$ follows from our assumption that the top quark 
Yukawa coupling is invariant under the flavor symmetries of the 
theory. There are several ways in which we may extend the $Z_{2}$ 
symmetry to the $H$, $F$, and $f$ fields that are consistent with the
interactions in (\ref{eq:whiggs}).  The simplest choice is:
\begin{center}
\begin{tabular}{cccccc}
$\overline{F}$ & $F$ & $f$ & $\varphi$ & $\varphi'$ & $H$ \\ \hline
      $-$       & $+$   &  $+$   &   $-$     &    $+$      &  $+$ 
\end{tabular}
\,\, .
\end{center}
It is interesting to note that the form of our superpotential interactions may
be guaranteed by other discrete symmetries.  One rather nice possibility is a
$Z_{4}$ symmetry, where the fields have the charges:
\begin{center}
\begin{tabular}{ccccccc}
$\overline{F}$ & $F$ & $f$ & $\varphi$ & $\varphi'$ & $H$ \\ \hline
      $i$       & $i$   &  $i$   &    $-$     &     $+$      &  $-$   
\end{tabular}
\,\, .
\end{center}
In this case, the $a$ vev breaks this $Z_{4}$ down to a $Z_{2}$ 
which is precisely the matter parity (or $R$-parity) needed to forbid
dangerous baryon- or lepton-number-violating interactions at the
renormalizable level.

We will proceed with our discussion assuming that the form of
our superpotential interactions are restricted by the
$Z_2$ symmetry introduced earlier.  All that remains is to 
specify the sector of the theory that is responsible for electroweak 
symmetry breaking (EWSB). This is considered in the next section.

\subsection{Electroweak Symmetry Breaking}

Perhaps the simplest solution for EWSB is achieved by assuming the
couplings
\begin{equation}
\Delta W_1 = \beta_{\varphi'} \varphi'_\mu \ F F \,\,\, ,
\label{eq:w1}
\end{equation}
\begin{equation}  
\Delta W_2 = \lambda_5 \varphi'_\mu H_u H_d - \frac{\lambda_6}{3} 
\varphi_\mu^{\prime 3} 
\,\,\, .
\label{eq:nmssm}
\end{equation}
The FN field $F$ in $\Delta W_1$ is neutral under standard model gauge
interactions.  
The field $\varphi'_\mu$ is even under the $Z_2$ symmetry like the
$\varphi'$ fields introduced earlier, though we assume that it has no
couplings to the other flavons $\varphi$ to simplify the discussion. 
Then, Eq.~(\ref{eq:nmssm}) is the usual superpotential of the next-to-minimal
supersymmetric standard model, except we assume that $\varphi'_\mu$ and $H$ 
have nontrivial flavor transformation properties.   This must be the
case if we are to prevent couplings $\chi H H$ or $\eta H H$.
The Higgs fields $H_u$ and $H_d$ acquire positive squared masses at 
the scale $\Lambda_{GM}$ from gauge mediation diagrams, so that 
$m^2_H(\Lambda_{GM}) \approx 350$ GeV. The contribution to the 
Higgs masses from the flavor sector is negligible given our choice of 
scales, $\Lambda_{\it flav} \sim \Lambda_{GM}/10$.  If we had chosen them 
to be comparable, we would have had an additional negative contribution to 
the $m^{2}_{H} \simeq - (3~\mbox{TeV}^{2})^{2}$ which is too large for 
correct electroweak symmetry breaking.
As the Higgs masses are run to lower energies, $m^2_{H_u}$ becomes negative 
due to the effect of the top quark Yukawa coupling and the squark 
squared masses.  However, the heaviness of the squarks in models 
with gauge-mediated SUSY breaking forces $m^2_{H_u}$ to become negative 
much more rapidly than it does in the MSSM, so that 
$m^2_{H_u}\approx -(500\mbox{GeV})^2$ at the weak scale. If $\varphi'_\mu$ 
has no soft supersymmetry breaking mass, then the simple extension of
the Higgs sector in Eq.~(\ref{eq:nmssm}) does not work phenomenologically:
there are always scalar and pseudoscalar states that are light enough
to be produced in $Z$ decay.  This problem led the authors of
Refs.~\cite{DN,DNS,DNNS} to consider much more complicated Higgs 
sectors. In our framework, the situation is somewhat better.
The coupling of $\varphi'_\mu$ to the flavor sector of the model through 
Eq.~(\ref{eq:w1}) leads to a two-loop negative mass squared 
$m^2_{\varphi'_\mu} \approx - (100 \mbox{ GeV})^2$, like the flavon 
fields $\varphi$.  The negative mass squared and the $\lambda_6
\varphi_\mu^{\prime 3}/3$ term force both $\langle \varphi'_\mu \rangle$
and $\langle F_{\varphi'_\mu} \rangle = \lambda_6 \varphi^{\prime 2}_\mu$
to become non-vanishing, and hence generate $\mu$ and $m_3^2$ parameters.
We studied the potential following from Eq.~(\ref{eq:nmssm}) numerically, 
and obtained local minima in which all the physical scalar and pseudoscalar 
states are sufficiently heavy.\footnote{A soft trilinear 
coupling $A \varphi'_\mu H_u H_d$ is generated at one loop, which pushes the 
lightest pseudoscalar mass above 10 GeV.  This is sufficient to
evade the bounds from astrophysics and cosmology, quarkonium decay,
and beam dump experiments.} However, this required a 
fine-tuning of the couplings $\lambda_5$ and $\lambda_6$.  Thus, 
the longstanding problem of generating the $\mu$ and $m_3^2$ parameters 
naturally in models with gauge-mediated supersymmetry breaking is not 
immediately resolved in our framework.  We considered other EWSB 
superpotentials, {\em e.g.}  those allowing $\varphi'_\mu$ to have couplings
to the other flavons of the form $\varphi^2 \varphi'_\mu$, but our 
conclusions remain unchanged.  Regardless of the details of the 
superpotential, we always found that a fine-tuning of parameters was 
necessary to compensate for the very large negative value 
of $m^2_{H_u}$.

Of course, it is possible that there is some explicit model of flavor 
compatible with our framework, that provides additional contributions 
to $\mu$, $m_{H}^{2}$ and $m_3^2$.  For example, in some model there 
may be FN fields with even matter parity that mix with Higgs fields.  
Such a mixing induces a positive mass squared to the Higgs bosons.  
Or, there may be higher dimension operators of the form $a (\langle 
\varphi \rangle/M_{F}) H_{u}H_{d}$, which generates
$\mu$ and $m_{3}^{2}$ at the desired orders of 
magnitude. Since this is a 
model-dependent question, we will not pursue this issue further here.

\subsection{R Axions}

The absense of dimensionful parameters in the theory we have
presented leads to an effective global $R$-symmetry under which
all fields transform with charge $2/3$.  Since our superpotential
is partitioned into ``sectors", which are relatively isolated
from each other, one may worry that there are separate and
potentially dangerous $R$-axions associated with each.  In this section,
we show that all the model-independent $R$-axions that are present
in our framework are phenomenologically harmless.  There could
be additional light scalar bosons that arise as a consequence of
accidental global symmetries in specific flavor models; we discuss
how these may be avoided in Section 4.1.

There are four approximate $R$ symmetries in our framework, corresponding
to each of the nearly decoupled sectors in which spontaneous
symmetry breaking occurs:  the supersymmetry breaking sector at the scale
$\Lambda_{SSB}$, the gauge mediation sector at $\Lambda_{GM}$, the
flavor sector at $\Lambda_{flav}$, and the EWSB-FSB sector at a
few hundred GeV.  The lines in Fig.~1 that connect the different
sectors explicitly break the independent U(1)$_R$ symmetries,
leaving one unbroken linear combination acting on all the sectors.
This corresponds to the non-anomalous $R$-symmetry in the
supersymmetry breaking sector, with all fields in the other sectors
transforming with charge $2/3$.  The remaining three linear
combinations are explicitly broken by the messenger U(1) gauge
interactions, the indirect coupling between the $a$ field and the
flavons through loops of FN fields in the flavor sector, and
the indirect coupling between the Higgs fields and $S$ through
loops of the $X$ field, in the gauge mediation sector.  Once the
U(1)$_R$ is spontaneously broken in the supersymmetry breaking 
sector, the low-energy effective theory beneath $\Lambda_{SSB}$
contains explicit U(1)$_R$ breaking parameters.  Below we
estimate the masses for the one true $R$-axion and the
three `would-be' $R$-axions separately.
  
The only true $R$-axion is the first linear combination described
above.  Since its decay constant is high, $F\sim 10^7$ GeV, all direct
search experiments are irrelevant. The only potential problem is its
possible contribution to the cooling of red giant stars.
However, we expect this $R$ axion to obtain a mass of order
$100$ MeV via the same mechanism which cancels the cosmological
constant \cite{BPR}, and thus it is astrophysically harmless.
The cosmological implications of this $R$-axion are less clear.
Since this issue is not specific to our framework, we will not
consider it further.

The would-be $R$-axion in the gauge mediation sector obtains a mass of 
order 10~TeV in the following manner.  Since global $R$-symmetry is 
spontaneously broken in the dynamical supersymmetry breaking sector, 
the messenger U(1) gaugino acquires a Majorana mass at the one-loop 
level.  We explicitly checked this point in the case of the 
SU(6)$\times$U(1) model.  Then a trilinear coupling $-A h' \xi'_{+} 
\xi'_{-} S'$ is generated at the two-loop order with $A \sim 
\Lambda_{SSB}/(16\pi)^{2} \sim \Lambda_{GM}/(16\pi)$.  This coupling 
explicitly breaks the global $R$-symmetry in the gauge mediation 
sector, and the $R$-axion aquires a mass of order $m_{A^0}^{2} \sim 
\langle A h' \xi'_{+} \xi'_{-} \rangle / \Lambda_{GM}^{2} \sim 
(10~\mbox{TeV})^{2}$ according to  Dashen's formula.  It is 
completely harmless given this large mass.

Similarly, the would-be $R$-axion in the flavor sector  obtains a 
mass via the analogous trilinear coupling 
$-A h \xi_{+} \xi_{-} \chi$.  While its mass 
is much smaller, $\sim 100$~GeV, it is 
still heavy enough to avoid all existing phenomenological constraints.  
Recall $\chi$ does not acquire a vev in the absence of the trilinear 
coupling, and hence $\langle \chi \rangle \sim A \sim 
\Lambda_{\it flav}/16\pi^{2}$ as discussed in Section \ref{subsec:flav}.  
Therefore the operator which explicitly breaks $R$-symmetry is 
suppressed.  Dashen's formula gives $m_{A^0}^{2} \sim \langle A 
\xi_{+}\xi_{-}\chi \rangle / \Lambda_{\it flav}^{2} \sim 
(100~\mbox{GeV})^{2}$.

Finally, there is a separate would-be $R$ axion in the flavon superpotential 
that gains a mass of order $10$ GeV as a consequence of the soft 
trilinear flavon interactions.  Trilinear couplings of the form $A 
\varphi^{2} \varphi'$ are generated at order $A \sim 1$~GeV through 
one-loop diagrams involving the FN fields and an insertion of their 
supersymmetry breaking bilinear mass term $\propto \langle 
F_{a}\rangle$.  As long as (at least) one of $\varphi'$ obtains a vev 
comparable to $\varphi$, the $R$-axion obtains a mass of order 
$m_{A^0}^{2} \sim \langle A \varphi^{2} \varphi' \rangle/\langle 
\varphi\rangle^{2} \sim (10~\mbox{GeV})^{2}$.  In Section \ref{subsec:flav}
we showed that the relative size of the two couplings $\gamma$ and $\delta$
determines whether 
or not $\varphi'$ acquires a vev.  If $\langle \varphi' \rangle = 0$
at lowest order for all $\varphi'$, then 
$\langle \varphi' \rangle$ is induced only through trilinear couplings, 
and the $R$-axion mass goes down to the 1~GeV level.  In this case, the 
quarkonium decay $\Upsilon \rightarrow A^0 \gamma$ excludes the model.  
On the other hand, there are no constraints from flavor-changing 
processes like $K^{0} \rightarrow \mbox{virtual}~A^0 \rightarrow 
\overline{K}^{0}$ because this $R$-axion couples to the overall 
$R$-charge 2/3 of each ordinary matter fields and its coupling is 
therefore flavor-blind.  Note also that the coupling of $A^0$ is axial and hence 
proportional to the fermion masses; this makes it impossible to find 
$A^0$ as an $s$-channel resonance at $e^{+} e^{-}$ experiments.  The 
beam dump experiments do not constrain axion-like fields above the GeV 
range.  Known astrophysical sources do not produce particles above the GeV 
range either.  Cosmology is also not likely to constrain such a 
particle because it decays relatively quickly as $A^0 \rightarrow 
b\bar{b}$ with a width much larger than a keV. We are not aware of any 
experimental constraints which exclude the existence of a scalar boson 
in the 10~GeV range which couples universally to all fermion axial 
currents.

\subsection{The Flavor Changing Problem} \label{subsec:fcp}

Now that we have outlined the important features of our model,
we return to the issue of flavor changing neutral currents,
and how they constrain our choice of scales.  In our framework, the 
ordinary squarks and sleptons receive four contributions to their
squared masses: 

({\em i})  A positive, flavor-blind contribution from
gauge-mediation diagrams, of order $\Lambda^2_{GM}/(16\pi^2)^2$.

({\em ii}) A negative, flavor-symmetric contribution from two-loop
diagrams like those in Figure~2, except with the $\varphi$ and $f$ 
lines interchanged, of order $-\Lambda^2_{\it flav}/(16\pi^2)^2$.  The term 
``flavor-symmetric'' refers to operators which respect the flavor 
symmetry of the model, without necessarily being flavor blind.
This can be the case, for 
instance, if the flavor symmetry is Abelian.

({\em iii}) A positive, flavor-dependent contribution due to the 
supersymmetry breaking scalar masses of the FN fields.  Note that the one-loop
subdiagram in Figure~2 will give the scalar $F$ and
$\overline{F}$ fields supersymmetry breaking squared masses, of order
$\Lambda^2_{\it flav}/(16 \pi^2)$.  This alters the (1,1) and (2,2) entries
of the scalar mass matrix in Eq.~(\ref{eq:negeig}), leading to
a flavor-dependent shift in the lightest eigenvalue of order 
$+\Lambda^2_{\it flav}/(16\pi^2) 
(\langle \varphi \rangle / M_F)^2$.

({\em iv})  A negative, flavor-dependent contribution as
in (3.7),due to the small
nonvanishing $F$ component of the field $a$. With
$\langle F_a \rangle
\approx \langle a \rangle^2/(16 \pi^2)$, this effect is of order 
$-\Lambda^2_{\it flav}/(16\pi^2)^2 (\langle \varphi \rangle / M_F)^2$.

The first contribution was estimated in Section~\ref{subsec:gauge},
assuming $\Lambda_{GM} = \langle F_S \rangle/\langle S \rangle = 
100$~TeV.  With $\Lambda_{\it flav} = 10$~TeV, we may estimate the remaining
contributions: 
\begin{center}
\begin{tabular}{cc}
({\em ii})  & $-(100$~GeV$)^2$ \\ 
({\em iii}) & $+(1000$~GeV$)^2 (\langle \varphi \rangle/M_F)^2$ \\ 
({\em iv})  & $-(100$~GeV$)^2 (\langle \varphi \rangle/M_F)^2$ 
\end{tabular} 
\end{center}
Notice that contribution ({\em ii}) is much smaller than contribution 
({\em i}), given the choice $\Lambda_{GM} = 10 \Lambda_{\it flav}$.  Thus, the
flavor-blind component of the squark and slepton masses is exactly what we 
would expect in the kind of scenario proposed by Dine and Nelson.  
If the flavor symmetry does not guarantee degeneracy of the squarks
(or sleptons) of the first two generations, then the flavor-symmetric
contributions ({\em ii}) can lead to flavor changing neutral current
effects.  In the quark sector, the constraints on $K$-$\overline{K}$
mixing are satisfied rather easily for 1~TeV squarks, so there
is no restriction on the flavor structure of the model.  However,
the lightness of the sleptons makes the situation in the lepton sector
more dangerous; the flavor-symmetric contributions ({\em ii}), may favor 
some flavor models over others.  Note that in models with a 
non-Abelian flavor
symmetry in which the first two generations transform as a doublet, there will
be no constraint on the contribution ({\em ii}).  The flavor-dependent 
contribution ({\em iii}) dominates over ({\em iv}), since the latter is 
suppressed by the the smallness of $\langle F_a \rangle$ in the FN
sector of our model. 
With $\Lambda_{\it flav}=10$ TeV, ({\em iii}) is marginally consistent with 
the bounds from flavor changing processes, assuming that the 
$\langle \varphi \rangle / \Lambda_{\it flav}$ are of order $10^{-1}$ in the 
quark sector, and a few $\times 10^{-2}$ in the lepton sector.  
Since ({\em iii}) 
scales as $\Lambda_{\it flav}^2$,  we would not be able to construct a viable 
model had we chosen $\Lambda_{\it flav}$ to be much larger than $10$ TeV.  
On the other hand, we will see in Section~4 that the exchange of the 
relatively light flavon fields are also marginally consistent with the 
bounds on FCNC processes, for $\Lambda_{\it flav}\approx 10$ TeV.  Thus, 
lowering $\Lambda_{\it flav}$ significantly is also phenomenologically 
unacceptable.  Our choice for $\Lambda_{\it flav}$ is a reasonable
compromise, given the constraints on flavor changing processes detailed
in Section~4.  

The constraints on the left-right mass matrices, on the other hand, 
are very weak in our framework.  The left-right masses
originate from the effective Yukawa couplings,
\begin{equation}
W_{\it eff} = \left( \frac{\varphi}{\Lambda_{\it flav}} \right)^n
      Q d H_d + \mbox{up-quark, leptons} \,\,\, .
\end{equation}
when one $\varphi$ field is set to its $F$ component, while
the remaining $\varphi$ fields and the Higgs field $H$ are all set
to their vevs.  Thus, the left-right mass terms are of the order
\begin{equation}
m_{LR}^2 \simeq m_f 
      \left(\frac{\langle F_{\varphi} \rangle}{\langle \varphi \rangle}
            \right) \,\,\, ,
\label{eq:mlr}
\end{equation}
where $m_f$ stands for the mass of a light quark or lepton.  
Therefore, the left-right mass terms are always proportional to the 
corresponding fermion masses, which is not necessarily true in 
the case of supergravity.  In the quark sector, all squarks are at 
1~TeV while the effective $A$-parameters are about 400~GeV or less.  
This is phenomenologically safe by itself. In the lepton secotr, equation
\ref{eq:mlr} may lead to a large mixing between smuons and selectrons,
and hence an unacceptably large 
$\mu \rightarrow e\gamma$ decay rate. Fortunately, this can
be avoided in a number of ways. For instance, if $\delta < 2\gamma$ 
(see discussions in Section \ref{subsec:flav}), $\varphi'$ does not 
acquire a vev and hence $F_{\varphi}$ vanishes identically.  A 
trilinear coupling among flavons induce $\langle \varphi' \rangle$ 
only at a higher order in $1/16\pi^{2}$.   In this case there is no further 
restriction on the flavor model.  On the other hand, an alignment or 
non-Abelian flavor symmetry can suppress the off-digonal entry in 
the left-right mass matrix in the basis where fermion masses are 
diagonal.

Finally, one may worry that operators involving the $F$-component 
of the $a$ field may contribute to left-right mass mixing at a more
dangerous level than the operators involving $F_\varphi$.%
The effective Yukawa operators of the form 
\begin{equation}
      W_{\it eff} = \left( \frac{\varphi}{M_{F}} \right)^{n} f f H
\end{equation}
generate trilinear couplings in the following way.  Since $M_{F} = 
\alpha \langle a\rangle$, they should be written more correctly as
\begin{equation}
      W_{\it eff} = \left( \frac{\varphi}{\alpha a} \right)^{n} f f H\, ,
\end{equation}
and an expansion of the chiral superfield $a$ around its vev as $a = 
a + \theta^{2} F_{a}$ generates trilinear couplings
\begin{equation}
      V_{\it eff} = \left( \frac{\varphi}{\alpha \langle a \rangle}
\right)^{n}
      n \left( \frac{\langle F_{a}\rangle}{\langle a \rangle} \right)
      \tilde{f} \tilde{f} H\,.
\end{equation}
Recall $\langle F_{a}\rangle / \langle a \rangle \sim \langle a 
\rangle/16\pi^{2} \sim 100$~GeV. If the power $n$ is different for 
different generations, the left-right mass terms generated from this 
operator cannot be simultaneously diagonalized with the fermion masses.  
This is not a problem in the quark sector, but could be serious in the 
lepton sector.  One way to avoid this problem is to have an alignment.  
Another way is to obtain the Yukawa couplings of the first two 
generations in any given Yukawa matrix from a set of operators in the 
high-energy theory that all involve the same powers of $\langle 
\varphi \rangle/M_{F}$; then these contributions to $m^2_{LR}$ will be 
diagonal in the fermion mass basis, and will no give
additional constraints.

\section{Phenomenology of a low flavor scale}
\setcounter{equation}{0}

In the previous section, we found that the flavor-blind, gauge-mediated
contribution to the scalar masses dominates over any flavor-dependent
splittings induced by the mixing of the light families with the FN fields. 
Thus, the super-GIM mechanism is effective and the usual supersymmetric
flavor problem ({\em i.e.} the large FCNCs effects generated by the exchange 
of sfermions in loops) is greatly reduced.   As a consequence, the usual
constraints placed on the flavor structure of the model are significantly
weakened in our framework.  This is in sharp contrast to the case of
gravity-mediated supersymmetry breaking, where the flavor symmetry 
must either guarantee a high degree of sfermion degeneracy or an alignment
between fermion and sfermion mass matrices to avoid the SUSY flavor problem. 

However, with the FN scale at $\sim 10$ TeV and flavon vevs and masses  
in the few hundred GeV range, we must consider new contributions to FCNCs
originating from the exchange of the physical states of the flavor sector. 
In this section, we describe the phenomenological constraints that 
FCNC processes impose on theories with low flavor scales.  Most of the new
flavor-violating effects are decribed by four-fermion operators, with
coefficients $C/M^2$. In Table~1, we present bounds on the coefficients $C$
assuming $M=M_F=10$ TeV, for a number of four-fermion operators that
contribute to rare processes.  We then estimate the coefficients $C$ for the
flavor-violating operators that may arise in our framework, and determine 
in what ways the bounds in Table~1 constrain the flavor structure of the
model.

\setcounter{footnote}{0}
There are four logical possibilities as flavor groups: combinations
of local or global, and discrete or continous.  In all cases, 
flavor-violating operators may be generated at two distinct scales: at 
$\sim 10$~TeV, where the FN fields $F$ and $\overline{F}$ are 
integrated out, and at a few hundred ~GeV where the flavon fields 
$\varphi$ are integrated out.  If there exist yet lighter degrees 
of freedom such as pseudo-Nambu-Goldston bosons, they may induce 
further flavor-changing operators.  We will discuss general constraints 
which apply to all cases first in the Section 4.1. By
itself, this discussion will present all the constraints relevant
to discrete flavor symmetries, except those relating to a possible domain
wall problem; we do not 
have anything to add on this point.\footnote{It may be worth  recalling 
that a global discrete symmetry could be anomalous.  Then the domain 
walls dissolve due to instanton effects and do not cause cosmological 
embarassments \cite{Preskill}.  On the other hand, a global discrete 
symmetry is probably spoiled by quantum gravitational effects.  
Still, it could well arise as an accidental low-energy symmetry 
especially when one considers a low flavor scale as in our framework.
Non-renormalizable operators suppressed by 1/$M_{pl}$ may also 
solve the domain wall problem even if the discrete symmetry is anomaly
free \cite{DN}}  
On the other hand, a broken continous flavor symmetry produces 
additional degrees of freedom (flavor gauge bosons in the  local case and 
``familons'' in the  global case) which induce new flavor-changing 
phenomena.  These two cases are discussed separately in Sections 
4.2 and 4.3.   
Most of the constraints are new as far as we know, though some were
briefly discussed in \cite{LNS2}.

\renewcommand{\topfraction}{0.9}
\renewcommand{\textfraction}{0.1}
\begin{table}[t]

\begin{center}
\begin{tabular}{||c|c|c||} \hline
Process & $O$ & $C<$ \\ \hline
   & $(\bar{d}^A \gamma^{\mu} L s_{A})(\bar{d}^B \gamma_{\mu} L s_{B})$
& $4\times 10^{-5}$ \\ \cline{2-3}
  &$(\bar{d}^{A} L s_{A})(\bar{d}^{B} L s_{B})$ & $6\times 10^{-6}$ \\
  \cline{2-3}
$\Delta m_K$  &$(\bar{d}^{A} L s_{B})(\bar{d}^{B} L s_{A})$ & $3\times
10^{-5}$
\\ \cline{2-3}
  &$(\bar{d}^{A} L s_{A})(\bar{d}^{B} R s_B)$ & $5 \times 10^{-6}$ \\
  \cline{2-3}
  &$(\bar{d}^{A} L s_{B})(\bar{d}^{B} R s_A)$ & $2 \times 10^{-5}$ \\ \hline
   & analogous to above & $4\times 10^{-4}$ \\ \cline{2-3}
  &  " & $6\times 10^{-4}$ \\ \cline{2-3}
$\Delta m_D,\Delta m_B$ &" & $3\times 10^{-3}$ \\ \cline{2-3}
  &" & $5\times 10^{-4}$ \\ \cline{2-3}
  &" & $4\times 10^{-3}$ \\ \hline
  & $(\bar{\mu} \gamma^{\mu} L e)(\bar{e} \gamma_{\mu} L
e)$
& $2 \times 10^{-3}$ \\ \cline{2-3}
$\mu \rightarrow 3 e$  & $(\bar{e} L \mu)(\bar{e} L  e)$& $9 \times 10^{-3}$ \\ \cline{2-3}
  & $(\bar{e} L \mu)(\bar{e} R  e)$& $7 \times 10^{-3}$ \\ \hline
$K_L \rightarrow \mu^+ \mu^-$ & $(\bar{d} \gamma^{\mu} L s)(\bar{\mu}
\gamma_{\mu} L \mu)$ &$4 \times 10^{-2}$
\\ \cline{2-3}
  &$(\bar{d} L s)(\bar{\mu} L \mu),(\bar{d} L s)(\bar{\mu} R \mu)$ & $4 \times 10^{-3}$ \\ \hline
$K_L \rightarrow \mu e$ & $(\bar{d} \gamma ^{\mu} L s)(\bar{\mu}
\gamma_{\mu} L e)$ &$4 \times 10^{-3}$
\\ \cline{2-3}
  &$(\bar{d} L s)(\bar{\mu} L e),(\bar{d} L s)(\bar{\mu} R e) $ & $3 \times 10^{-4}$ \\ \hline
$K_L \rightarrow e^+ e^-$ & $(\bar{d} \gamma^{\mu} L s)(\bar{e}
\gamma_{\mu}L e)$ &$6 \times 10^{-1}$
\\ \cline{2-3}
  &$(\bar{d} L s)(\bar{e} L e),(\bar{d} L s)(\bar{e} R e)$ & $3 \times 10^{-4}$ \\ \hline
\end{tabular}
\end{center}
\caption{Constraints on the four fermion operators ${C\over {M^2}} O$
from various rare processes, with $M=$ 10 TeV. $A,B$ are color indicies. Other
relevant operators can be obtained from the above by charge 
conjugation.  This table then exhausts all possible four fermion
operators contributing to these flavor changing processes. 
$L$ and $R$ are chiral projection operators for 
left-handed and right-handed fields, respectively.
We have used $(m_s + m_d) =$ 160 MeV. The bounds on 
$\Delta m_i$ scale as (160 MeV)$^2$/$f_i^2 B_i$, where
$i=K,D,B$ and $f_i,B_i$ are the relevant decay and bag constants respectively.}
\end{table} 

\subsection{General constraints from FCNC}

The existence of FN fields at 10~TeV scale and flavons at a few 
100~GeV scale can induce new flavor-changing processes.  The 
phenomenological constraints are identical for discrete or continous 
flavor symmetries, except those induced by flavor gauge bosons or 
familons.  They will be discussed separately in the next subsections.  
We discuss constraints common to all possible flavor groups in this 
subsection.  
        
\subsubsection{Flavor-violating operators induced at FN scale.} 

The $\overline{F} \varphi f$ couplings lead to diagrams with $f$ fields
on the external lines, and the FN and flavon fields in loops.  
Flavor-violating operators, like those listed in Table~1, are induced in the 
low-energy theory when the FN and $\varphi$ fields are integrated out.  
All superpotential couplings which we generically refer to as $h$ are 
$O(1)$, so these operators are 
only suppressed by loop factors and the mass of the FN fields 
$M_F \sim 10$ TeV.  In our numerical estimates below, we
set all the $\overline{F} \varphi f$ Yukawa couplings to $1$, and
assume that the multiplicity of the particles running around internal
loops is also $1$. We consider flavor violating processes in the lepton, 
quark and mixed lepton-quark sectors separately. 

\noindent $\bullet$ Lepton Sector
  
\begin{figure}[t]
	\centerline{\psfig{file=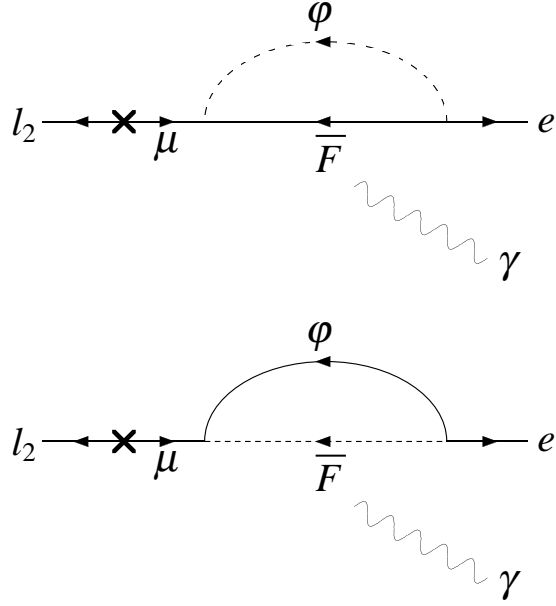,width=0.5\textwidth}} 
	\caption{Feynman diagrams contributing to $\mu\rightarrow e\gamma$.}
	\protect\label{mu-e-gamma}
\end{figure}

$\mu \rightarrow e \gamma$: 
The amplitude for $\mu \rightarrow e \gamma$ comes from the diagrams of
Fig.~3.  Crucially, this amplitude, like all magnetic transitions, 
vanishes in the supersymmetric limit.  For $M_{F}$ = 10 TeV, we find
\begin{equation}
B(\mu \rightarrow e \gamma) \sim  10^{-10} h^4 
\left[f(m_{\varphi_R}^2/{M_F^2}) 
+ f(m_{\varphi_I}^2/{M_F^2}) - 2 f(m_{\varphi_f}^{2}/{M_s^2})\right]^2
\label{eq:meg}
\end{equation}
where the $m_{\varphi_{R,I}}$ are the masses of the real and imaginary 
components of the scalar part of $\varphi$, $m_{\varphi_f}$ is the mass of 
the fermionic part of $\varphi$,  $M_F$ is the mass of the FN fermion, $M_s$ 
the mass of the FN scalar, and 
\begin{equation}
f(x) = {{2 + 3x - 6x^2 + x^3 + 6x \log(x)}\over{12 (x - 1)^4}}.
\end{equation}
Since ${m_{\varphi}\over M_F} \sim 10^{-2}$, the quantity in square brackets in
Eq.~(\ref{eq:meg}) is highly suppressed; we find that it is always numerically
much smaller than $\sim .001$.  Thus, $B(\mu \rightarrow e \gamma)
\mbox{\raisebox{-1.0ex}{$\stackrel{\textstyle ~<~}{\textstyle \sim}$}}
10^{-16}$, well beneath the current bound $B(\mu \rightarrow e \gamma) < 5
\times 10^{-11}$. We conclude that the $\mu \rightarrow e \gamma$ operator 
is not dangerous, even if it is allowed in the flavor-symmetric limit. 

\begin{figure}[t]
	\centerline{\psfig{file=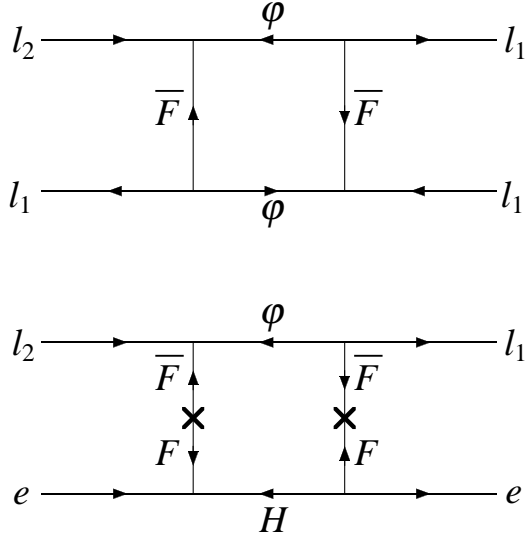,width=0.5\textwidth}}
	\caption{Superdiagrams contributing to $\mu\rightarrow eee$.}
	\protect\label{lepton-box}
\end{figure}

$\mu \rightarrow 3 e$: 
If allowed by the flavor symmetry, box diagrams with internal FN and flavon
fields can generate four-fermion operators that contribute to
$\mu \rightarrow 3 e$.  Even if these operators are forbidden in the flavor
symmetric limit, they may be generated after we rotate a flavor-symmetric
operator to the mass eigenstate basis.  The diagram of Fig.~4a, (the flavor
symmetric version of which necessarily exist) generates the interaction
\begin{equation}
{h^4 \over{32 \pi^2 M^2}} 
(\bar{\mu} \gamma^{\mu} L e)(\bar{e} \gamma_{\mu} L e)
\end{equation}
Even if this operator is allowed in the flavor symmetric limit,
the coefficient $C \sim 3 \times 10^{-3}$ is in borderline agreement
with the constraint given in Table~1.

The diagram of Fig.~4b, on the other hand, does not neccesarily exist. 
It can only be generated if there is a direct Yukawa coupling between the
Higgs, FN and lepton fields of the first two generations. This box diagram
diverges in the infrared, and is cut off by the mass $m_{\varphi}$. 
From this diagram we generate the operator 
\begin{equation}
{h^4 \over {8 \pi^2 M^2}}(1 - \mbox{log}(M/m_{\varphi})) 
(\bar{e} L \mu)(\bar{e} R  e)
\,\,\, .
\end{equation}
With $m_{\varphi}/M_F \sim 10^{-2}$, the coefficient of this operator is
$C \sim 4\times 10^{-2}$, which is bigger than the bound in Table~1 by 
a factor of $\sim 7$. This is not neccesarily a disaster, but it is certainly
safer to forbid the diagram 4b in the flavor symmetric limit. If the above
operator is generated through mixing with an angle $\sim \sqrt{m_e/m_{\mu}}
\sim 0.1$ it is no longer dangerous phenomenologically. 

\noindent $\bullet$ Quark Sector

$b \rightarrow s \gamma$: The rate is completely negligible, for the same
reasons given in our discussion of $\mu \rightarrow e \gamma$.

\begin{figure}[t]
        \centerline{\psfig{file=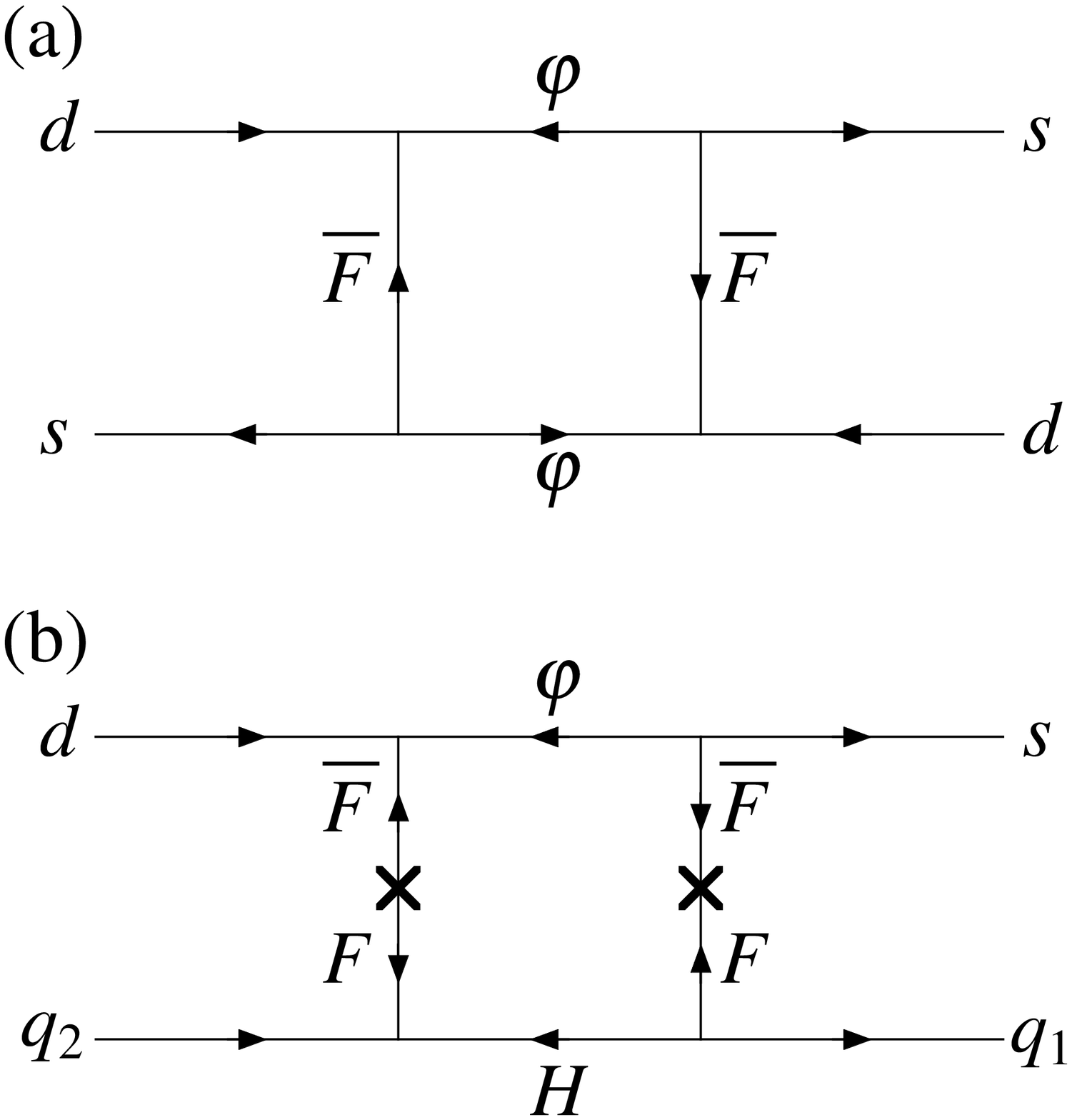,width=0.5\textwidth}
        \psfig{file=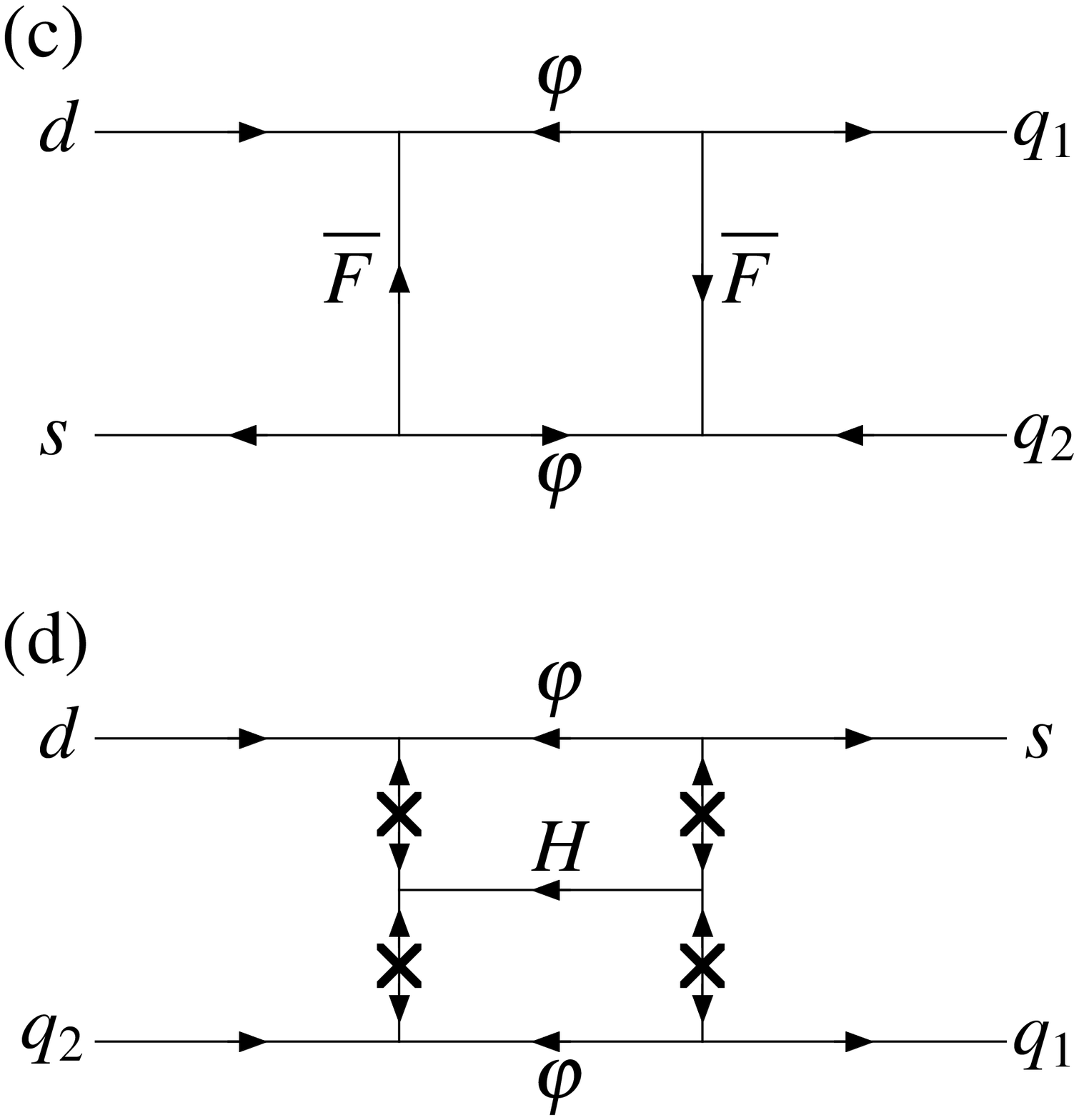,width=0.5\textwidth}}
	\caption{Superdiagrams contributing to neutral meson mixing.}
	\protect\label{quark-box}
\end{figure}

$\Delta m_K$: The diagram of Fig.~5a generates the interaction 
\begin{equation}
{h^4 \over{32 \pi^2 M^2}} 
(\bar{d}^A \gamma^{\mu} L s_{A})(\bar{d}^B \gamma_{\mu} L s_{B})
\,\,\, .
\end{equation} 
The coefficient $C \sim 3\times 10^{-3}$ is $\sim 75$ times bigger than the
bound in Table~1.  Thus, this diagram must be forbidden in the flavor
symmetric limit. With a mixing suppression of $\sim \lambda^2$, the
coefficient is only 3 times bigger than the bound. However, it is
easy to compensate for this factor by choosing all the Yukawa couplings 
in the diagram to be $1/2$ instead of $1$. 

The diagram in Fig~5b only exists if there are flavons coupling to both 
left- and right-handed superfields. If such couplings exist, we generate
the operator
\begin{equation}
{h^4 \over{16 \pi^2 M^2}} 
(\bar{d}^{A} L s_{B})(\bar{d}^{B} L s_{A})
\,\,\, ,
\end{equation}
after appropriate Fierz transformations.  The coefficient
$C \sim 6\times 10^{-3}$ is $\sim 200$ times bigger than the bound in Table~1,
and we conclude that this diagram must be forbidden in the flavor symmetric
limit.   If we include a mixing supression $\sim \lambda^2$, the coefficient
is still too large by a factor of 10. In this case, however, we can  
reduce the amplitude sufficiently by choosing all the Yukawa couplings in the
diagram to be $h \sim 1/2$ instead of $1$. 

The diagram in Fig~5c is by far the most dangerous one we will encounter. It
only exists if there is a direct Yukawa coupling between down or strange, FN
and Higgs fields.  As in the case of $\mu \rightarrow 3 e$, this box diagram
is enhanced by an infrared divergence.  We obtain the operator 
\begin{equation}
{h^4 \over {8 \pi^2 M^2}}(1 - \mbox{log}(M_F/m_{\varphi})) 
(\bar{d}^{A} L s_{A})(\bar{d}^{B} R s_{B})
\,\,\, .
\label{eq:disast}
\end{equation}
If we take $m_{\varphi}/M_F \sim \lambda^2$, the coefficient
$C\sim 2\times 10^{-2}$, is $\sim 3000$ times larger than the
corresponding bound in Table~1; even with $\lambda^2$ suppression, it is
still $\sim 100$ times too big.  Notice that we cannot reduce the magnitude
of this operator by choosing smaller couplings in the box diagram. The
product of the two couplings on the left-hand side of the diagram 
with ${{\langle \varphi \rangle}\over M_F}$ gives us an element of a quark 
Yukawa matrix; thus, we can only reduce the couplings in the box diagram if 
we increase ${{\langle \varphi \rangle}\over M_F}$.  Note also that this
diagram is 
particularly worrisome in a theory with large $\tan\beta$.  Recall that the
negative squared masses for the flavon fields were naturally of order ${1
\over {(16 \pi^2)^2}} M_F^2$, and thus ${{\langle \varphi \rangle}\over M_F}$
fell in the range $\lambda^2 - \lambda^3$.  However, for large $\tan\beta$,
the strange Yukawa coupling is itself of order $\lambda^2 - \lambda^3$, and 
we therefore require a direct coupling between the strange, Higgs and a FN
field. This is precisely the situation that gives us the disastrous
contribution to $\Delta m_K$ in Eq.~(\ref{eq:disast}). 

Finally, even when there is no direct Yukawa coupling between
down or strange, Higgs and FN fields, we can have a two-loop contribution to
$K-\bar{K}$ mixing as shown in Fig~5d. We generate exactly the same effective
operator as in Eq.~(\ref{eq:disast}), and we find for $m_{\varphi} \sim
\lambda^2 M_F$, a coefficient $C \sim 3\times 10^{-5}$, which is 5 times bigger
than the bound.  This can easily be avoided either by choosing slightly
smaller couplings in the diagram, or by forbidding the operator in the flavor
symmetric limit. 

It is important to point out that our conclusions remain unchanged if we
include CP violating phases of ${\cal O}(1)$ in the theory.  In this
case, the constraint from $\epsilon_K$ is $\sim 10$ times stronger than
the one from $\Delta m_K$ that we have presented here.  However, we
have already concluded from the FCNC considerations that the
corresponding $\Delta S = 2$ operators must be forbidden in the flavor
symmetry limit.  In a model where this is the case, CP violation does
not provide us with any additional generic constraints.

$\Delta m_D, \Delta m_B$: The only significant constraint
from
either of these processes comes from the analogue of the dangerous diagram 5c
above.  We conclude that this diagram should be forbidden
in the flavor symmetric limit; if it is induced through mixing from a flavor
symmetric operator, then its effects are on the borderline from
experimental constraints.

\noindent $\bullet$ Mixed quark-lepton sector

If some flavons couple to both the quark and lepton sectors, they can give a
significant contribution to $K_L \rightarrow \ell^+ \ell^-$ via box 
diagrams analogous to those in Figs.~5a-d.  We have already learned that we 
must forbid direct couplings between the down/strange, Higgs and FN fields, 
so we do not consider the analogue of Fig.~5c: this diagram would give too 
large an amplitude for $K \rightarrow \mu e,ee$ by a factor $\sim 200$
if it were generated in the flavor symmetric limit. All the remaining 
diagrams are on the borderline even if they are present in the flavor 
symmetric limit.

\subsubsection{Flavor violating operators induced at the flavor symmetry
breaking  
scale.} 

These operators arise from tree-level exchange of physical flavons. 
Recall that we generate higher dimension operators of the form  
\begin{equation}
\left({\varphi \over M_F}\right)^{n} f_i f_j^c H  
\label{eq:gym}
\end{equation}
after integrating out the FN fields. Here the $f_i$ are ordinary fields of
the first two generations.  When the $\varphi$ fields aquire vev's, the 
operator in (\ref{eq:gym}) gives us an element of the corresponding 
Yukawa coupling matrix. If we now set all but one of the flavon fields
and the Higgs field to their vevs, we generate a Yukawa coupling between the
light fermions and flavons:
\begin{equation}
\left({\langle \varphi \rangle \over M_F}\right)^{n-1} 
{\langle H \rangle \over M} 
\varphi f_i f_j^c\\
\sim {m_{ij} \over {\langle \varphi \rangle}} f_i f_j^c \varphi \,\,\, .
\label{eq:ffphi}
\end{equation}
We do not show an exact equality in Eq.~(\ref{eq:ffphi}) since, in general,
the mass matrix element $m_{ij}$ receives contributions from several different
operators of the form (\ref{eq:gym}), with different $O(1)$ coefficients. Of
course, it may be the case that in specific models, the flavon couplings 
will become flavor diagonal when we rotate to the fermion mass basis.
Generically, however, this will not be the case, and we will obtain 
flavor-violating four-fermion operators when we integrate out the physical
flavons at tree level. Let us  write $m_{ij} = \max(m_i,m_j) \Theta_{ij}$.
Then, fixing the mass of the flavons coupling only to the quark sector at
$\sim \lambda^2 \times 10$ TeV $\sim 400$ GeV and the mass of the flavons
coupling to the leptons or leptons and quarks at $\sim 100$
GeV,\footnote{See Section \ref{subsec:flav} for these estimates.} 
we present the strongest constraints on the magnitudes  of the
$\Theta_{ij}$ in Table~2. In all cases, the $\Theta_{ij}$ can be as large or
larger than the corresponding CKM matrix element, and thus physical flavon
exchange does not give us significant constraints.

\begin{table}
\begin{center}
\begin{tabular}{||c|c||} \hline
Process & constraint \\ \hline
$\mu \rightarrow 3 e$ & $\Theta_{\mu e} < 130$ \\ \hline
$\Delta m_K$ & $\Theta_{ds} < .2$ \\ \hline
$\Delta m_D$ & $\Theta_{uc} < .3$ \\ \hline
$\Delta m_B$ & $\Theta_{bd} < .09$ \\ \hline
$K_L \rightarrow \mu^+ \mu^-$ & $\Theta_{ds} < .2$\\ \hline
\end{tabular}
\caption{Constraints on $\Theta_{ij}$ from exchange of physical flavons.
The mass and vev of the flavons coupling to quarks alone are taken to be
at 400 GeV, all others at 100 GeV.} 
\end{center}
\end{table}
\subsubsection{Pseudo-Nambu-Goldstone bosons} 

The above analysis assumes that all the physical flavons get masses of 
order the vev of the flavon field.  However, given many flavons and 
the restriction to renormalizable superpotentials, it is often the 
case that there are approximate, accidental continuous symmetries of 
the tree-level potential, producing pseudo-Nambu-Goldstone bosons 
(PNGB) which pick 
up a mass only at loop level $\sim \langle \varphi \rangle / 4 
\pi$ which may be as small as 10~GeV. These PNGBs can have 
flavor-violating couplings, and (especially for the PNGBs coupling 
to the first two generations) mediate disastrously large FCNC.  Thus, a 
specific model must ensure that most of the PNGBs receive a mass directly at 
tree level.  This puts a constraint on the flavor symmetry and flavon 
particle content.  The absence of accidental global symmetries must be 
checked in each explicit model of flavor.

Even when there is no accidental global symmetry which results in 
PNGB with flavor-dependent coupling, there is always one 
model-independent PNGB.  Since the superpotential is purely trilinear, it 
neccessarily has a tree level $R$ symmetry under which all fields have 
$R$ charge $2/3$ and the negative squared masses for the flavon fields 
do not break this $R$ symmetry.  We discussed this particular 
``model-independent $R$-axion'' in Section 3.6 and showed it does not have 
flavor-changing interaction and is phenomenologically harmless as long 
as it is heavier than 10~GeV.

\subsubsection{Summary of General Constraints}  

Let us summarize the major constraints that emerged from our analysis. We
have found that the flavor symmetry must forbid all $K$-$\bar{K}$ mixing
operators in the flavor symmetric limit. Furthermore, there can be no direct
Yukawa coupling of the $FfH$ type between down/strange, Higgs and FN fields;
this in particular 
causes great difficulty for a scenario with large tan$\beta$. 
Once these constraints are satisfied, CP violation does not put any
further significant constraints even with $O(1)$ phases.
Notice that the
constraints on the flavor group we have found are quite different from the usual ones 
needed to guarantee sfermion degeneracy. For instance, an $SU(2)$ flavor
symmetry with the first two generation fields in a doublet is sufficient to
guarantee sfermion degeneracy in the flavor symmetric limit; however the
dangerous operator $(\epsilon^{jk} \overline{d}_i R q_k)
(\epsilon^{kl}\overline{q}_k R d_l)$
(with $i,j,k,l$ flavor $SU(2)$ indices) gives large $K$-$\bar{K}$ mixing
while being  
completely flavor symmetric. Similarly, $U(1)$'s (or discrete subgroups) can 
be used to forbid all $K$-$\bar{K}$ mixing operators in the flavor 
symmetric limit but cannot in themselves guarantee sfermion  degeneracy.  
Also, a specific model must sufficiently break any accidental global 
symmetries which give rise to light PNGBs. These new 
constraints differ from the ones we usually encounter in supergravity 
scenarios, and suggest new avenues for flavor model building. 

\subsection{Continuous, global flavor symmetries}

In addition to the constraints presented in the previous subsection,
continuous flavor symmetries lead to other, often problematic, contributions
to flavor changing processes.  In the case of global flavor symmetries,
we must contend with the Nambu-Goldstone bosons (``familons'') that arise
when the flavor group is spontaneously broken. If some NG
bosons have flavor-violating couplings (which is certainly the case if the
group has a non-Abelian component), FCNC constraints, from processes like $K
\rightarrow \pi +$ familon and $\mu \rightarrow e+$ familon, push the flavor
symmetry breaking scale above $\sim 10^{11}$ GeV. However, if the flavor
group has only $U(1)$ factors, the familons may have purely diagonal couplings
in the flavor basis. If the first two generation fields have different charges
under some $U(1)$ factor, the alignment between flavor and mass eigenstate
bases must be very precise ($\sim  1\mbox{ TeV}/10^{11}\mbox{ GeV} =
10^{-8}$) to avoid FCNC constraints, which implies that a mechanism for
perfect alignment is required. For instance, we can imagine that all left
handed fields in the theory have the same charge, whereas the right-handed
fields have different charges. Since the only absolute requirement we have is
for the existence of non-trivial rotations on the left handed fields to
generate the CKM matrix, the flavor and mass bases may be  exactly aligned for
the right-handed fields. Then, the rotation in going to the mass eigenstate
basis for the left-handed fields does not induce any off-diagonal familon
coupling (since the left handed charges are generation blind), while there is
no rotation on, and hence no off-diagonal familon coupling to the right handed
fields. While this sort of idea is not excluded, it is clear that continuous
global flavor groups are strongly constrained by the requirement of purely
diagonal familon couplings. 

\subsection{Continuous, local flavor symmetries}
In the case of a gauged flavor symmetry, a new contribution to flavor
violating processes comes from the exchange of massive flavor gauge bosons 
at the $\sim 100$ GeV scale.  This source of FCNCs will place significant
restrictions on the form of the flavor group and symmetry breaking sector,
as we will see below.  In addition, when a gauged flavor symmetry is 
spontaneously broken, degeneracy between the first and second generation
sfermions may be spoiled by flavor-dependent $D$ terms in the scalar
potential \cite{hitoshi}.  We consider both issues below.
   
\noindent $\bullet$ Flavor gauge boson exchange 

Let us examine the effects of gauge boson
exchange in the down quark sector. Suppose that the left-handed Weyl spinor
quarks $q_L$ (grouped into a three-vector in generation space) transform
under a 
representation $T^{a}_L$ of the flavor group while the right-handed down
quarks $d_R$ transform under $T^{a}_{d_R}$. In the flavor basis (denoted
by primes), the flavor current is given by 
\begin{equation}
{J^{a}}_{\mu} = \bar{d^{\prime}_L} \bar{\gamma}_{\mu} 
T^a_L d^{\prime}_L + \bar{d^{\prime}_R} \bar{\gamma}_{\mu} T^a_{d_R}
d^{\prime}_R \,\,\, .
\end{equation}
After integrating out the massive flavor gauge bosons, we generate the
following four-fermion operators:
\begin{equation}
\sum_a {g^2 \over {M_a^2}} J^{a \mu} J^{a}_{\mu}\\
=\sum_a {g^2 \over {M_a^2}} 
(\bar{d^{\prime}_L} \bar{\gamma}_{\mu} T^{\prime a}_L d^{\prime}_L 
+ \bar{d^{\prime}_R} \bar{\gamma}_{\mu} T^{\prime a}_{d_R} d^{\prime}_R )^2
\end{equation}
In the second expression we have rotated to the mass basis $q = U_q
q^{\prime}$, $d^c = U_{d^c} d^{c \prime}$, and the $T^{a \prime}$ are the
flavor group generators in the mass basis, $T^{a \prime} = U T^a U^{\dagger}$.
Suppose that the pattern of flavor symmetry breaking is such that the $M_a$
are different from each other. If the flavor group is Abelian and there is
precise alignment between flavor and mass eigenstates, all the operators above
are flavor diagonal. However, if there is no alignment, the $M_a$ must be
pushed above $\sim 1000$ TeV to avoid FCNC constraints (in particular,
from $\Delta m_K$). Thus, for an Abelian flavor symmetry broken below the TeV
scale, the alignment between mass and flavor bases must be precise
better than
$O({{1\mbox{TeV}}/{1000\mbox{TeV}}}) = 10^{-3}$. For non-Abelian flavor
groups, a given $T^{a \prime}$ will have off-diagonal flavor-violating
elements. When the $M_a$ are all different, there is no hope that summing over
$a$ will yield a flavor conserving result, once again forcing the $M_a$ to
above 1000 TeV. Thus, if the flavor group is non-Abelian and broken at the TeV
scale, some mechanism must guarantee that the flavor gauge bosons (especially
coupling to the first two generations) have identical mass at least at tree
level. For this to happen it seems neccessary to have some  accidental
``custodial" symmetry analogous to the one which forces $\rho = 1$ at tree
level in the standard model. We have not succeeded in finding a model of
flavor which simultaneously guarantees sufficient flavor gauge
boson degeneracy and produces the Cabibbo angle, but this remains 
an interesting direction to explore.
 
\noindent $\bullet$ $D$ term splitting

In Ref.~\cite{hitoshi}, it was shown that flavor $D$ terms can split
the first two generation sfermion masses by an amount independent of
the flavor gauge couplings, if flavor symmetry breaking occurs in the
supersymmetric limit.  In this case, the splittings cannot be made small 
by reducing the flavor gauge coupling.  This is also the case in
supergravity scenarios that generate $\langle \varphi \rangle /M_F$
along flat directions after supersymmetry is broken.  In our framework,
however, the situation is different.  The flavor sector potential
has a stable minimum as the flavor gauge coupling is taken to zero.
Thus, the flavor $D$ terms can be made arbitarily small.

As an example, consider an $SU(2)$ theory with a doublet flavon
$\varphi_a$ and a 
triplet field $\Sigma^{ab}$, with superpotential $W=\lambda \varphi_a \varphi_b
\Sigma^{ab}$. Suppose that only the doublets talk to the FN fields and hence
get a negative mass squared at two loops. The crucial point is that, even if
the gauge coupling is put to zero, the potential has a stable minimum; in the
$g \rightarrow 0$ limit the superpotential part of the potential is minimized
by putting $\Sigma^{ab} = 0$, and we have 
\begin{equation}
V=|\lambda|^2 (\varphi^{\dagger} \varphi)^2 - m^2 \varphi^{\dagger} \varphi
\end{equation}
and without loss of generality we can choose $\langle \varphi \rangle =
({m\over 
{\sqrt{2} |\lambda|}}, 0)$. For sufficiently small $g$, the $D$ term
contribution is a small perturbation to the above potential. In particular,
the vev of the flavon $D$ term is $\sim m^2$, and so the induced $D$ term
splitting between first two generation sfermions is $\sim g^2 m^2$. Now, for
squark masses of $\sim 1$ TeV, $K$-$\bar{K}$ mixing constrains $\theta_c
{(\tilde{m}_{d}^2 - \tilde{m}_{s}^2) \over {\tilde{m}^2}} < .01$. Thus, 
for $m \sim 500$  GeV we must have $g^2 < 1/5$, making this gauge coupling
moderately larger than $e$. Given that many fields transform under this flavor
$SU(2)$, it could be that the $SU(2)$ coupling is non-asymptotically free, in
which case a small value for the coupling could be naturally explained. 

In the lepton sector, it is difficult to ensure that the $D$ terms don't spoil
slepton degeneracy without running into trouble with a very light flavor gauge
boson coupling to leptons. The reason is that the $D$ terms splitting has the
form $\delta m_{\widetilde{\ell}} ^2 \sim g^2 \varphi^{\dagger} T^a
\varphi \sim 
M^2$ where $M$ is the mass of the flavor gauge boson. On the other hand, for
mixing in the slepton masses of $m_{\widetilde{\ell}}$ $\sim$ 100 GeV and
slepton mixing of $\sim \sqrt{{m_e}\over{m_{\mu}}}$, the constraint from $\mu
\rightarrow e \gamma$ demands that $\delta m_{\widetilde{\ell}}
^2/m_{\widetilde{\ell}}^2 <$ .01, putting the flavor gauge boson at a mass
less than about $30$ GeV. Thus, in order to have gauged flavor symmetries in
the lepton sector, we must have some mechanism either to cancel the unwanted
$D$ term splitting between selectron and smuon masses or to guarantee the
absense of the slepton mixing in the first two generations.

\section{Conclusions}
\setcounter{equation}{0}

Understanding the origins of supersymmetry breaking (SSB) 
and flavor symmetry breaking (FSB) are two of the greatest challenges for
theories with weak-scale supersymmetry. 
Much of the structure of the theory is dictated
by the mechanisms and scales for these symmetry breakings. In particular, the
degree to which the soft supersymmetry breaking operators contain information
about flavor depends on the relative size of the flavor scale, $M_F$, and the
messenger scale for supersymmetry breaking $M_{mess}$. In supergravity
theories
with $M_F < M_{mess} \approx M_{Pl}$, the interactions of the flavor scale
leave
an imprint on the soft operators; while in theories of 
gauge-mediated supersymmetry breaking, with $M_{mess} < M_F$, they do not. 

In this paper we have studied a framework in which $M_{mess}$ and $M_F$
are comparable because they have a common origin -- the scale of
dynamical supersymmetry breaking, $\Lambda_{SSB}$.  This scheme allows a
unified view of FSB and EWSB -- indeed the FSB vevs $\langle \varphi
\rangle$ are comparable to the EWSB vevs $\langle H \rangle$. This
unification is illustrated in Figure 1, which shows the Froggatt-Nielsen
sector as being the flavor analogue of the gauge mediation messenger
sector. Indeed these sectors bear more than just a passing resemblance:
both contain heavy vector generations of matter. Fundamentally, they are
distinguished only by whether these heavy vector generations have large
supersymmetry breaking contributions to their masses. We have given
explicit models for the messenger sector, (\ref{eq:ndsup}) + (3.9), and
for the Froggatt-Nielsen sector (\ref{eq:Wxi}) +
(\ref{eq:DWxi}). These models are both variants of a basic model which
has two phases, with the vacuum choice dependent on the values of the
dimensionless couplings. The sectors are chosen to be in opposite
phases, so that the heavy vector generations feel
supersymmetry breaking strongly in the gauge mediation sector but only mildly in
the flavor sector. These models, while certainly not unique, illustrate
our scheme and explicitly show how the flavon fields acquire negative
squared masses  triggering FSB, in a way which is analogous but not
identical to the triggering of EWSB.

The interactions which feed supersymmetry breaking to the superpartners are 
phenomenologically very important. Supergravity mediation has been the most
studied case, and mediation by the known gauge interactions is the only other
case that has received significant attention. Our scheme does have mediation
via the known gauge interactions, but in additonal there is mediation via the
superpotential interactions of the Froggatt-Nielsen sector. This provides a
new origin for contributions to the soft operators; in particular it
provides the dominant contribution to the soft trilinear A terms, and
important
flavor dependent contributions to the scalar mass matrices.

We believe that the unified scheme for FSB and EWSB introduced in this paper, 
and summarized in Figure 1, provides 
significant motivation for studying a new class of models for flavor.
This new class of models, although based on the old ideas of Froggatt and
Nielsen, are substantially different from the theories which have been
constructed up to now. This is partly because the messenger sector provides a
dominant, flavor-independent, contribution to the squark and slepton masses,
and
partly because flavon vevs of order the weak scale introduce
several new flavor-changing constraints. 

The construction of any 
Froggatt-Nielsen model requires a choice for the flavor group, $G_f$, 
and for the $G_f$ transformations of the flavons $\varphi$, the heavy vector 
generations $F$ and the light generations $f$. The class of models which is
consistent with the framework of this paper has these choices
severely restricted by the following six constraints:

1) $G_f$ is prefered to be discrete  -- a continuous global $G_f$ gives unacceptable
   familons, while a continuous gauged $G_f$ has additional flavor
   dependent scalar mass contributions from $D^2$ terms.

2) The size of the flavon vevs is given by
   $\langle \varphi \rangle /M_F \approx 1/16 \pi^2 \approx \lambda^2
   \mbox{or} \lambda^3$, where $\lambda = 0.22$. This hierarchy is determined
   by the order in perturbation theory at which the flavon vevs are generated,
   and the requirement that all dimensionless couplings be of order unity. The
   top quark Yukawa interaction must be $G_f$ allowed.

3) Sufficient trilinear flavon interactions must be $G_f$ allowed  so that 
   there are no accidental U(1) flavor groups which are spontaneously broken. 
   Since there must be several different flavon fields, this is a very
powerful
   constraint on the theory.

4) Box diagrams involving internal heavy vector generations and flavons must
   not generate $K_L - K_S$ mixing in the $G_f$ symmetric limit. For example,
   this means that the 12 or 21 entries of the down Yukawa matrix may not be
   generated at linear order in $\varphi/M_F$. This excludes the case of very 
   large $\tan \beta$ where the $b$ Yukawa coupling is of order unity, since
in
   that case these entries are expected to be linear in $\varphi$ in order to
   generate the Cabibbo angle. The only way to avoid this conclusion is if the
   Cabibbo angle is generated from the up sector, in which case $D-\bar{D}$
   mixing is predicted at the level of the present experimental limit.

5) From the previous point it follows that the $b$ and $\tau$ Yukawa couplings
   should arise at linear order in $\varphi$. This means that $\tan \beta$ is
   expected to be low, less than about 3. The mixing $V_{cb}$ should arise
   at order $\varphi$ in the up sector, or $\varphi^2$ in the
   down sector. 
   Entries of the light $2 \times 2$ block of the Yukawa
   matrices should be at most of order $\varphi^2$.

6) To prevent dangerous non-universal $A$ terms, all non-zero entries of the 
   light $2 \times 2$ block of the down and lepton Yukawa matrices must be of 
   order $\varphi^2$.

Each of the above constraints is very significant, and none of them applies to
Froggatt-Nielsen models with a high flavor scale and supergravity mediated
supersymmetry breaking.
Furthermore, these constraints are completely
independent of the particular models chosen to give masses to the heavy vector
generations in the gauge mediation and Froggatt-Nielsen sectors; 
there may be additional model-dependent constraints.
For example, the explicit flavor sector of Section 3 which led to masses
for the heavy 
vector generations of the Froggatt-Nielsen sector was based on a $Z_2$
symmetry. The flavons separated into two categories: $\varphi$ which have
$\bar{F} \varphi f$ type couplings and $\varphi'$, which do not. The trilinear
interactions necessary to prevent accidental flavor $U(1)$s then take the form
$\varphi'^3 + \varphi' \varphi^2$. Finally there may be further constraints which
arise from the mechanism used to generate effective $\mu$ and $m_3^2$ terms.
In
Section 3.5 this was accomplished by an interaction of the form
$\varphi_\mu 
H_u H_d$, implying that at least one of the Higgs fields must transform 
non-trivially under the flavor group.

\begin{center}               
{\bf Acknowledgments} 
\end{center}
This work was supported in part by the Director, Office of 
Energy Research, Office of High Energy and Nuclear Physics, Division of 
High Energy Physics of the U.S. Department of Energy under Contract 
DE-AC03-76SF00098 and in part by the National Science Foundation under 
grant PHY-95-14797.  The work of NA-H was supported by NSERC.

\appendix
\section{Two-Loop Integrals}
\setcounter{equation}{0}

In this appendix, we evaluate the diagrams in Figure~2 using the difference
of propagators given in (\ref{eq:propdiff}):
\[
i \left[ p^2 - \left(\begin{array}{cc} 2g^2 & -2g\sqrt{2\lambda g} \\
                    -2g\sqrt{2\lambda g} & 4 g \lambda \end{array} \right)
\frac{m^2}{h^2}\right]^{-1} \;\;\;\;\;\;\;\;\;\;\;\;\;
\]\begin{equation} \;\;\;\;\;\;\;\;\;\;\;\;\;
-i \left[ p^2 - \left(\begin{array}{cc} 2(g^2+h^2) & -2g\sqrt{2\lambda g} \\
                    -2g\sqrt{2\lambda g} & 4 g \lambda \end{array} \right)
\frac{m^2}{h^2}\right]^{-1}             
\end{equation}
Since these diagrams only involve modification of the $a$ propagator, we
need the $(2,2)$ component of the expression above:
\begin{equation}
\frac{-i (16 \lambda g^3 m^6)}
{p^2[h^2p^2-2gm^2(g+2\lambda)][h^2p^4-2m^2(g^2+h^2+2g\lambda)p^2 
+8 g\lambda m^4]}
\end{equation}
The sum of the diagrams in Figure~2 then gives us
\[
\int \frac{d^4p}{(2\pi)^4}
\frac{-i (8 \alpha^2 \beta^2 g^3 \lambda m^6)}{[h^2p^2-2gm^2(g+2\lambda)]
[h^2p^4-2m^2(g^2+h^2+2g\lambda)p^2
+8 g\lambda m^4]} 
\]\begin{equation}
\times \int \frac{d^4l}{(2\pi)^4} \frac{l^2+M^2}{l^2(l^2-M^2)^2[(p-l)^2-M^2]}
\label{eq:momint}
\end{equation}
where $M=\alpha \langle a \rangle$ is the Froggatt-Nielsen mass scale.
Note that diagrams (B), (C) and (D) in Figure~2 are individually 
ultraviolet divergent, but their sum is finite.  The $d^4l$ integral in 
Eq.~(\ref{eq:momint}) can be done analytically by conventional  methods.
To evaluate the remaining $p$ integral, we first go to Euclidean 
space, and do the trivial angular integration.  What remains is a 
one-dimensional integral in the Euclidean radial coordinate $p_E$:
\begin{equation}
\frac{i}{(16\pi^2)^2} (16 \alpha^2 \beta^2 g^3 \lambda m^6)
\int_0^\infty dp_E p_E^3 F(p^2_E/M^2) D_1(p^2_E) D_2(p^2_E)
\label{eq:odint}
\end{equation}
where 
\begin{equation}
D_1(p^2_E) = \frac{1}{M^2[h^2p_E^2+2gm^2(g+2\lambda)]}  \,\,\, ,
\end{equation}
\begin{equation}
D_2(p^2_E) = \frac{1}{h^2p_E^4+2m^2(g^2+h^2+2g\lambda)p_E^2
+8 g\lambda m^4} \,\,\, ,
\end{equation}
and
\begin{equation}
F(x) = \frac{-4-2x}{\sqrt{x(4 + x)}}
\tanh^{-1} \sqrt{\frac{x}{4 + x}} + \frac{x + 1}{x}
\ln (1 + x).
\end{equation}
The function $F(x)$, plotted in Figure~6, is positive definite.
We have chosen $g$ and $\lambda$ to be real and positive without loss of
generality, by making suitable field phase
rotations. Therefore the entire integrand of 
Eq.~(\ref{eq:odint}) is also positive definite.  If we  use a
different phase convention, the vevs and mass spectrum change
accordingly but the final sign of the mass squared remains the same.  We
conclude that the 
mass squared generated in the flavon potential is negative. The $dp_E$ 
integral can be evaluated numerically, and the results are consistent with 
the order of magnitude estimate given in the text.

\begin{figure}[t]
	\centerline{\psfig{file=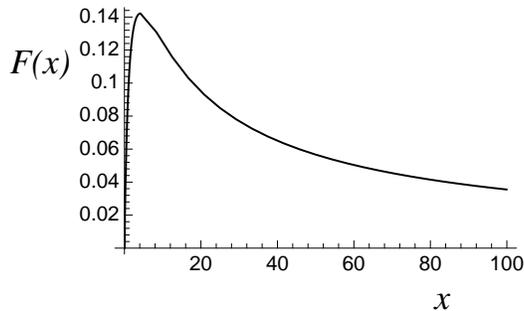,width=0.5\textwidth}}
	\caption{The function $F(x)$.}
	\protect\label{fig6}
\end{figure}

We have shown that the two-loop diagrams in Figure 2 generate finite
negative definite squared masses for the flavons.  One also obtains
apparently ultraviolet-divergent contributions at third or higher loop
orders in perturbation theory.  However, the ultraviolet divergence is
cutoff at $\Lambda_{SSB}$ because the supersymmetry is restored above
this scale, and one needs to take into account that the supersymmetry
breaking mass parameter $m^2$ vanishes above $\Lambda_{SSB}$ into
account.  Therefore higher order divergent contributions are suppressed
compared to the two-loop finite ones by a factor of
\begin{equation}
\sim \frac{1}{16\pi^2} \ln \frac{\Lambda_{SSB}^2}{m^2}
\simeq 0.09 \, ,
\end{equation}
and hence can be neglected.

\end{document}